\documentclass[twocolumn,twocolappendix,trackchanges]{aastex63}
\usepackage{lineno}
\usepackage{threeparttable}
\usepackage{appendix}
\usepackage{natbib}
\usepackage{color}
\usepackage{mathptmx}
\usepackage{amsmath}
\usepackage{url}
\usepackage{multirow}
\usepackage{verbatim}
\usepackage[utf8]{inputenc}
\usepackage{bm}

\newcommand{\msun}{\,\rm M_\odot}
\newcommand{\zsun}{\,\rm Z_\odot}

\newcommand{\kms}{{\rm km\;s}^{-1}}

\definecolor{ejcol}{rgb}{0,0.8,0}

\defcitealias{Shin20}{Paper I}
\defcitealias{Lee21}{Paper II}

\graphicspath{{./}}

\begin{document}

\shorttitle{DMDG Formation in a Mini-bullet Collision}
\shortauthors{Lee et al.}

\title{Multiple Beads-on-a-string: Dark Matter-Deficient Galaxy Formation in a Mini-bullet Satellite-satellite Galaxy Collision}

\correspondingauthor{Joohyun Lee}
\email{joohyun.lee@austin.utexas.edu}

\author[0000-0001-8593-8222]{Joohyun Lee}
\affiliation{Department of Astronomy and Texas Center for Cosmology and Astroparticle Physics, The University of Texas at Austin, Austin, TX 78712-1083, USA}

\author[0000-0002-4639-5285]{Eun-jin Shin}
\affiliation{Institute of Astronomy, University of Cambridge, Madingley Road, Cambridge CB3 0HA, UK}
\affiliation{Kavli Institute for Cosmology, University of Cambridge, Madingley Road, Cambridge CB3 0HA, UK}

\author[0000-0003-4464-1160]{Ji-hoon Kim}
\affiliation{Center for Theoretical Physics, Department of Physics and Astronomy, Seoul National University, Seoul 08826, Korea}
\affiliation{Seoul National University Astronomy Research Center, Seoul 08826, Korea}

\author[0000-0002-0410-3045]{Paul R. Shapiro}
\affiliation{Department of Astronomy and Texas Center for Cosmology and Astroparticle Physics, The University of Texas at Austin, Austin, TX 78712-1083, USA}

\author[0009-0002-3230-8205]{Eunwoo Chung}
\affiliation{Center for Theoretical Physics, Department of Physics and Astronomy, Seoul National University, Seoul 08826, Korea}

\begin{abstract}

Dark matter-deficient galaxies (DMDGs) discovered in the survey of ultra-diffuse galaxies (UDGs), in apparent conflict with standard CDM, may be produced by high-velocity galaxy-galaxy collisions, the {\it{Mini-bullet}} scenario.   
Recent observations of an aligned trail of $7-11$ UDGs near NGC1052, including DMDGs DF2 and DF4, suggesting a common formation event, $\sim8.9\pm1.5$ Gyr ago, provide a test.
Hydro/N-body simulations, supplemented by galaxy orbit integrations, demonstrate that satellite-satellite collisions outside the host-galaxy virial radius can reproduce the observed UDGs in the NGC1052 group.
A trail of $\sim10$ DMDGs is shown to form, including two massive ones that replicate the observed motions 
of DF2 and DF4.
The linear relation, $\bm{v}=A\bm{x}+\bm{v}_{0}$, conjectured previously to relate positions ($\bm{x}$) and velocities ($\bm{v}$) of the aligned DMDGs as a signature of the collision event, is approximately obeyed, but individual DMDGs can deviate significantly from it. 
The progenitors whose collision spawned the trail of DMDGs survive the collision without, themselves, becoming DMDGs. 
We predict one progenitor is located at the end of the trail, testable by observing the difference between its stars, formed pre-collision, from those of the DMDGs, formed post-collision.
By contrast, stellar ages and metallicities of the DMDGs are nearly identical.
We further offer a hint that the tidal field of host NGC1052 may contribute to making DMDGs diffuse.
$\Lambda$CDM simulation in a 100 cMpc box finds our required initial conditions $\sim10$ times at $z<3$. 
These results indicate current observations are consistent with the {\it{Mini-bullet}} scenario.

\end{abstract}


\vspace{-2mm}

\section{Introduction} \label{sec:intro}

In recent years, several galaxies have been observed to contain a lower amount of dark matter than predicted by galaxy formation theory in the standard Cold Dark Matter (``CDM'') model. The latter posits that halo formation occurs in the pressure-free, collisionless dark matter prior to the gravitational infall of the baryonic component. On very small mass scales, below the baryonic Jeans mass of the pregalactic medium, gas pressure in the baryons can resist gravity, making the baryon mass fractions of the lowest-mass halos below the cosmic mean baryon-to-dark matter density ratio.  For objects well above this baryonic Jeans-filter scale, the infall of baryons is supersonic and pressure forces are unimportant, so the baryons collapse along with the dark matter, and the baryon mass fraction inside virialized halos is close to that cosmic mean density ratio.  When gaseous baryons are heated by feedback processes inside (e.g. SNe, AGNs) and/or outside (e.g. reionization) of the halo to which they would have been bound, pressure forces can suppress their infall or reverse it, resulting in a baryon-to-dark ratio well below the cosmic mean.  However, in all these cases, the halos that form are dark matter-dominated.   
It was notable, therefore, when  \citet{vD18a, vD19} reported the existence of two ultra-diffuse galaxies (UDGs), NGC1052-DF2 and NGC1052-DF4 (hereafter DF2 and DF4, respectively), that are located in close proximity to the massive elliptical galaxy NGC1052 and exhibit a deficiency in dark matter, rather than the dark matter-dominance described above\footnote{UDGs are defined to be dwarf galaxies with effective half-light radius $R_{\rm eff}\ge1.5$ kpc and surface brightness $\mu(g,0) > 24\;{\rm mag}\;{\rm arcsec}^{-2}$ \citep{vD15}}.
Subsequently, dark matter-deficient galaxies (DMDGs) have been identified across various environments and mass scales. 
These include the local group and isolated low-mass galaxies \citep{Guo20}, a distant low-mass galaxy \citep{Mancera-Pina22}, and even a massive early-type galaxy in a cluster environment \citep{Comeron23}.

The formation model that explains DF2 and DF4's dark matter deficiency should also address their exceptionally luminous globular cluster (GC) population \citep{vD18b, vD19}.
To account for both phenomena at the same time, \citet{Silk19} proposed a ``Mini-Bullet (cluster)'' event or a ``Bullet dwarf'' scenario, in which a high-velocity ($\gtrsim$300 $\kms$) collision of low-mass (dwarf) galaxies dissociates collisionless dark matter from baryons. As the name suggests, this scenario was inspired by the famous example of separation of dark matter and baryons observed in the Bullet Cluster, which has been explained by the collision of two cluster-scale halos at a high velocity, greater than either of their virial velocities, in which the collisionless nature of CDM and stars allows those components of each colliding halo to pass through each other, while the baryonic intracluster gas in each is prevented from doing so by its fluid behavior, resulting in a bow shock seen in X-ray emission \citep{Tucker98, Liang00, Markevitch02, Clowe06}.
In the analogous {\it Mini-Bullet} collision, strong shock compression is induced, which in turn triggers star formation and the formation of massive star clusters. 
The large mass and narrow mass range of the observed star clusters are explained in this model by large gas surface densities that lead to a large lower limit to the initial cluster mass function while large galactic shear limits their mass range from above \citep{Trujillo-Gomez21}.
The galaxy collision produces these necessary conditions by strong radiative shocks that compress the gas and large-scale motions.

The separation of dark matter and baryons on various scales has been extensively studied on various scales, ranging from the GC scale \citep{Kim18, Ma20, Madau20} to the galaxy cluster scale \citep{Volker07, Milosavljevic07, Mastropietro08, McDonald22}, within the framework of the $\Lambda$CDM cosmology, using theoretical modeling and simulations.
With regard to the DMDGs, we previously demonstrated this \emph{Mini-Bullet} scenario using idealized galaxy collision simulations in \citet[][hereafter Paper I]{Shin20},
where we used two different N-body hydrodynamics codes with distinct numerical schemes to constrain the collision parameter space.
Furthermore, we showed that massive star cluster formation is indeed triggered by high-velocity galaxy collision and that the star cluster properties from the simulation are roughly in line with the observations \citep[][hereafter Paper II]{Lee21}.

The \emph{Mini-Bullet} model is not the only one advanced so far to explain the DMDGs. One of the most frequently studied alternative mechanisms for the formation of these unique systems is tidal interaction that transfers dark matter to the more massive system (tidal stripping) \citep{Ogiya18, Maccio21, Jackson21, Ogiya22a, Moreno22, Montero-Dorta23, Mitrasinovic23, Katayama23}.
In the case of DF2 and DF4, however, although there are measurements of their tidal distortion, the results do not require that the galaxies' dark matter was removed by tidal interaction \citep{Keim22}.
\citet{Muller19} also argued that there is no sign of stellar streams induced by tidal interaction near DF2 and DF4, which is also claimed by \citet{Montes21} with respect to DF2.
On the other hand, \citet{Montes20} claimed that DF4 is undergoing tidal disruption.
In any case, the tidal interaction scenario cannot explain the exceptionally bright GC population as a natural outcome of the scenario.

Another scenario is the tidal dwarf galaxy (``TDG'') formation mechanism, in which DMDGs formed from efficiently cooled and fragmented gas after it was ejected during a strong tidal encounter with a disk galaxy \citep{Recchi07, Duc14, vD19, Haslbauer19, Fensch19}.
This mechanism, too, has, so far, not been shown to involve the simultaneous formation of bright GCs, however.

Another idea, suggested by \citet{Trujillo-Gomez22}, explains the dark matter deficiency as a consequence of the formation of those bright GCs and their back-reaction on the gaseous galactic baryons, which in turn modified the dark matter distribution.
They argued that powerful stellar feedback from massive GC populations can induce a gravitationally-coupled expansion of the dark matter content, reducing its contribution to the dynamical mass of the galaxy \citep[also see][for similar work on the response of dark matter to gas ejection]{Li23}.

Recently, \citet{vD22} presented a new clue that supports the ``Mini-bullet'' scenario in the case of DF2 and DF4 by measuring their line-of-sight velocities. 
The authors conclude that both UDGs were formed from a single event that occurred about 8 Gyr ago, likely a high-velocity galaxy collision, which has been shown to be capable of producing the observed lack of dark matter and bright GC populations \citepalias{Shin20, Lee21}.
Using the HST observation of DF2 and DF4 and a catalog of low-surface brightness galaxies in the NGC1052 group studied by \citet{Roman21}, the authors identified an alignment of $7-11$ UDGs as a possible ``trail of DMDGs'' \citep[see Figure 1 in][]{vD22}.
This argument is further substantiated by follow-up observation performed by the same group, which revealed that the GCs of DF2 and DF4 have the same color \citep{vD22b}.
Additionally, age measurements of the GCs and stellar bodies by \citet{Fensch19} (DF2 only) and \citet{Buzzo22} (DF2 and DF4) yielded consistent results of $\sim 8$ Gyr for the age of the GCs and stellar bodies.

More recently, \citet{Buzzo23} studied the large-scale structure of GCs in the NGC1052 group and found that the GC distribution is consistent with scenarios involving a single galaxy-galaxy interaction event and subsequent coeval formation of the GCs and the DMDGs, which includes the tidal dwarf galaxy scenario and the {\textit {Mini-bullet}} scenario.
To distinguish the two scenarios, a possible ``smoking gun'' signature of the {\textit {Mini-bullet}} scenario was suggested by \citet{Gannon23}, a linear relationship between the line-of-sight velocities of the aligned DMDGs produced by the collision and their distance from their common point of origin in the collision; the further a galaxy was from this point, the larger must its launch velocity have been to reach that distance in the same elapsed time.
The authors noted that the known velocities and projected distances from NGC1052 of DF2 and DF4 were consistent with such a linear relationship. To test this further, \citet{Gannon23} used spectroscopy to measure the stellar age, stellar metallicity, and line-of-sight velocity of one more DMDG in the trail, NGC1052-DF9 (DF9).  
They concluded that the age and metallicity of the galaxy are similar to those of DF2 and DF4, consistent with the {\textit {Mini-bullet}} scenario, but the observed line-of-sight velocity of DF9 deviates significantly from the expected linear relationship.

In response to the observations, \citet{Ogiya22b} claimed the {\textit {Mini-bullet}} scenario may face challenges due to the strong tidal forces exerted by the host galaxy, NGC1052, which can strip GCs from the formed DMDGs shortly after their formation.
This argument is supported by two main factors.
Firstly, the spatial distribution of the observed GCs in DMDGs DF2 and DF4 is extended, and taking into account their orbital decay due to dynamical friction, their formation epoch-distribution should have been even more extended than their current distribution \citep{Dutta19, Dutta20}.
Secondly, the galaxy-galaxy collision occurs at or within the virial radius of NGC1052.
The authors argue that the combined effect of the extended distribution of GCs and the strong tidal field exerted by NGC1052 makes the {\textit {Mini-bullet}} scenario implausible.

In this paper, we shall explore the ability of the {\textit {Mini-bullet}} scenario to account for the enigmatic characteristics of the UDGs in the NGC1052 group, their dark matter deficiency, and alignment.
Our investigation will be carried out through a series of gravitohydrodynamic simulations and galaxy orbit integrations using the {\sc enzo} code and the {\tt Rebound} code, respectively. 
The initial conditions of the simulations will be designed to match the observed physical properties, including stellar masses and kinematics, and alignment of the NGC1052 group UDGs.
Our results will demonstrate that appropriate initial structural and orbital parameters of the colliding satellite progenitor galaxies can produce a ``trail of DMDGs'' that includes two massive DMDGs with $M_{\star} > 10^{8} \msun$ corresponding to DF2 and DF4, whose motions agree with the observed values.
We will show that while the positions and velocities of the DMDGs on the trail generally follow a linear relationship, as previously suggested to be a signature of their collision origin,
there can be deviations in the positions and velocities of individual DMDGs from that simple relation.
We will find that the stellar ages and metallicities of the DMDGs are nearly identical, but we will also examine the scatter in their values.
We will compare the simulated DMDGs with observed UDGs and discuss which physical processes need to 
be taken into account.
We will also quantify the occurrence of such {\textit {Mini-bullet}} events in the Universe using a large simulation of galaxy formation from cosmological initial conditions, TNG100-1.

This paper is structured as follows.
In Section \ref{sec:2}, we describe our effort to constrain the initial conditions by using idealized galaxy-galaxy collision simulations and backward (i.e. time-reversed) orbit integration.
Section \ref{sec:3.1} presents the simulation results, including the stellar masses and orbits (positions and velocities) of the DMDGs that formed.
In Section \ref{sec:3.2}, we discuss the stellar properties, stellar metallicities, ages, and sizes of the product DMDGs.
We compare these results with previous observational and theoretical work in Section \ref{sec:4.1}.
Section \ref{sec:4.2} discusses the statistical likelihood of the {\textit {Mini-bullet}} satellite-satellite galaxy collision in a large cosmological simulation {\sc IllustrisTNG}.
Our summary and conclusions are presented in Section \ref{sec:5}.

\section{Simulations} \label{sec:2}

We present a three-step methodology aimed at aligning hydrodynamic simulations with the observational findings of \citet{vD22}, focusing on the formation of multiple DMDGs through a single {\textit {Mini-bullet}} collision between two progenitor galaxies orbiting around the massive host halo of NGC1052.
Our primary objectives are to match {\it (\romannumeral 1)} the stellar mass of DF2 and DF4, the two most massive DMDGs among the NGC1052 group UDGs, {\it (\romannumeral 2)} the line-of-sight velocity difference of DF2 and DF4, {\it (\romannumeral 3)} the positions of DF2 and DF4, and {\it (\romannumeral 4)} the number of resultant DMDGs.

In the first step, we conduct idealized galaxy-galaxy collision simulation experiments similar to what we have done in \citetalias{Shin20} and \citetalias{Lee21} to explore the parameter space of progenitor galaxy properties, such as the dark matter halo mass ($M_{\rm DM}$), gas mass ($M_{\rm gas}$), gas distribution, and collision configuration, including relative collision velocity ($v_{\rm col}$) and pericentric distance ($r_{\rm min}$), that can produce $\sim 10$ aligned DMDGs after the collision. 
As the second step, utilizing the information obtained from the previous step regarding collision configurations capable of generating aligned DMDGs and their associated positions and velocities, we conduct backward orbit integrations to determine the initial conditions for the hydrodynamic simulations based on the observation of the NGC1052 group UDGs.
Finally, in the third step, we perform hydrodynamic simulations using the established initial conditions to examine the feasibility of the {\textit {Mini-bullet}} scenario in producing {\textit {a trail of DMDGs}} and predict their characteristics. 
In the following sections, we elaborate on each step in detail.

\subsection{Confining parameter space---idealized galaxy-galaxy collision simulations} \label{sec:2.1}

\begin{deluxetable*}{cccccccccc}[ht!]
    \tablenum{1}
    \vspace{-1mm}
    \tablecaption{
    A suite of 10 pc-resolution idealized dwarf galaxy-dwarf galaxy collision pair simulations listed with their initial configurations.
    \label{tab:1}
    }
    \vspace{-1mm}
    \tablewidth{0pt}
    
    \tablehead{
        \colhead{Run name} & \colhead{$v_{\rm col}$} & \colhead{$r_{\rm min}$} & \colhead{$R_{\rm s,\,gas}$} & \colhead{$M_{\rm total}$} & \colhead{$f_{\rm gas}$} & \colhead{$M_{\rm \star, \, DMDG, \, max}$} & \colhead{$\Delta v_{\rm DMDG, \, max2}$} & \colhead{$N_{\rm DMDG}$} & \colhead{$t_{\rm end}$}
    \\[-2mm]
        \colhead{} & \colhead{($\kms$)} & \colhead{(kpc)} & \colhead{(kpc)} & \colhead{($10^{10}\msun$)} & \colhead{} & \colhead{($10^{8}\msun$)} & \colhead{($\kms$)} & & \colhead{(Gyr)}
    \\[-2mm]
        \colhead{(1)} & \colhead{(2)} & \colhead{(3)} & \colhead{(4)} & \colhead{(5)} & \colhead{(6)} & \colhead{(7)} & \colhead{(8)} & \colhead{(9)} & \colhead{(10)}
    }
    
    \startdata
    {\tt 1} & 300 & 2 & 2 & 1.95 & 0.12 & 3.7 & 123 & 8 & 0.6\\
    {\tt 2} & 400 & 2 & 2 & 1.95 & 0.12 & 2.8 & 202 & 7 & 0.7\\
    {\tt 3} & 500 & 2 & 2 & 1.95 & 0.12 & 0.61 & 278 & 7 & 0.7\\
    {\tt 4} & 600 & 2 & 2 & 1.95 & 0.12 & 1.1 & 306 & 7 & 0.9\\
    \enddata
    
    \tablecomments{
        (1) run name,
        (2) relative velocity of the two progenitors at 60 kpc distance,
        (3) pericentric distance (i.e. distance at closest approach), 
        (4) scale radius of the PIS gas density profile,
        (5) the total mass of a progenitor,
        (6) gas fraction $f_{\rm gas} = M_{\rm gas} / M_{\rm total}$, 
        (7) stellar mass of the most massive DMDG formed, 
        (8) the largest relative velocity difference between the DMDGs, 
        (9) the number of formed DMDGs, 
        (10) time since the pericentric approach of the two progenitor galaxies when we end the simulation.}
    \vspace{-8mm}
\end{deluxetable*}

\begin{figure*}[t]
    \centering
    \includegraphics[width=\textwidth]{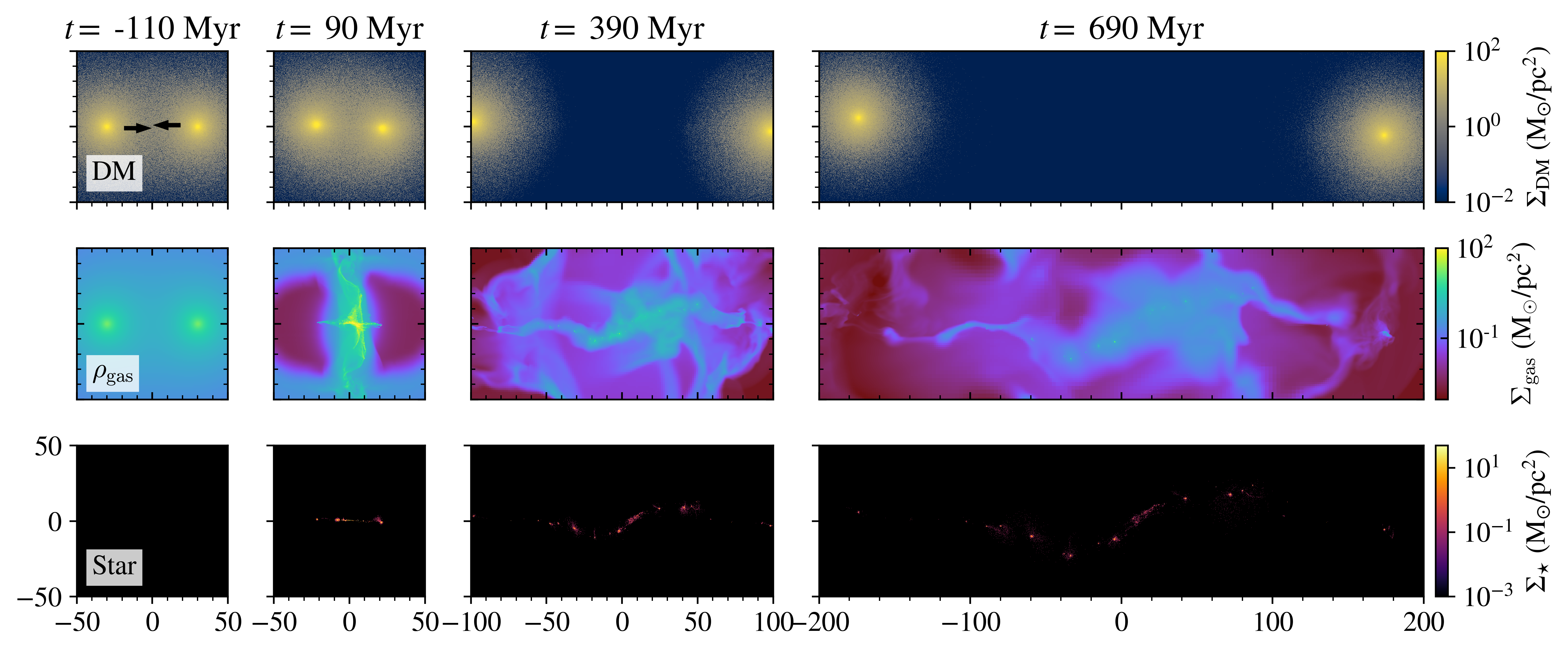}
    \caption{
        \textbf{\emph{Illustrative Mini-Bullet collision, before and after.}} 
        Snapshots of the time-history of the {\sc enzo} simulation of an idealized collision of two identical gas-rich dwarf galaxies, each with $M_{\rm total}=1.89\times10^{10}\msun$, 
        with a collision velocity of 500 $\kms$ (with black arrows indicating the progenitors' moving directions; Run {\tt 3} in Table \ref{tab:1}),
        at $t = -110$ (initial time-slice of simulation),$\;90,\;390,\;690$ Myr.
        $t = 0$ is set to the moment when the two dwarf galaxies are at pericenter (i.e. closest approach).
        Surface densities of dark matter (top row), gas  (middle row), and only those stars that formed after the start of the simulation (110 Myrs before the orbits of the colliding haloes reached pericenter) (bottom row) are presented.
        All projections are conducted in 100 kpc-depth layers.
        After the galaxy collision, seven DMDGs are formed and survive ($t = 690$ Myr; fourth column).
        \label{fig:1}
        }
\end{figure*}

\begin{figure*}[t]
    \centering
    \includegraphics[width=0.95\textwidth]{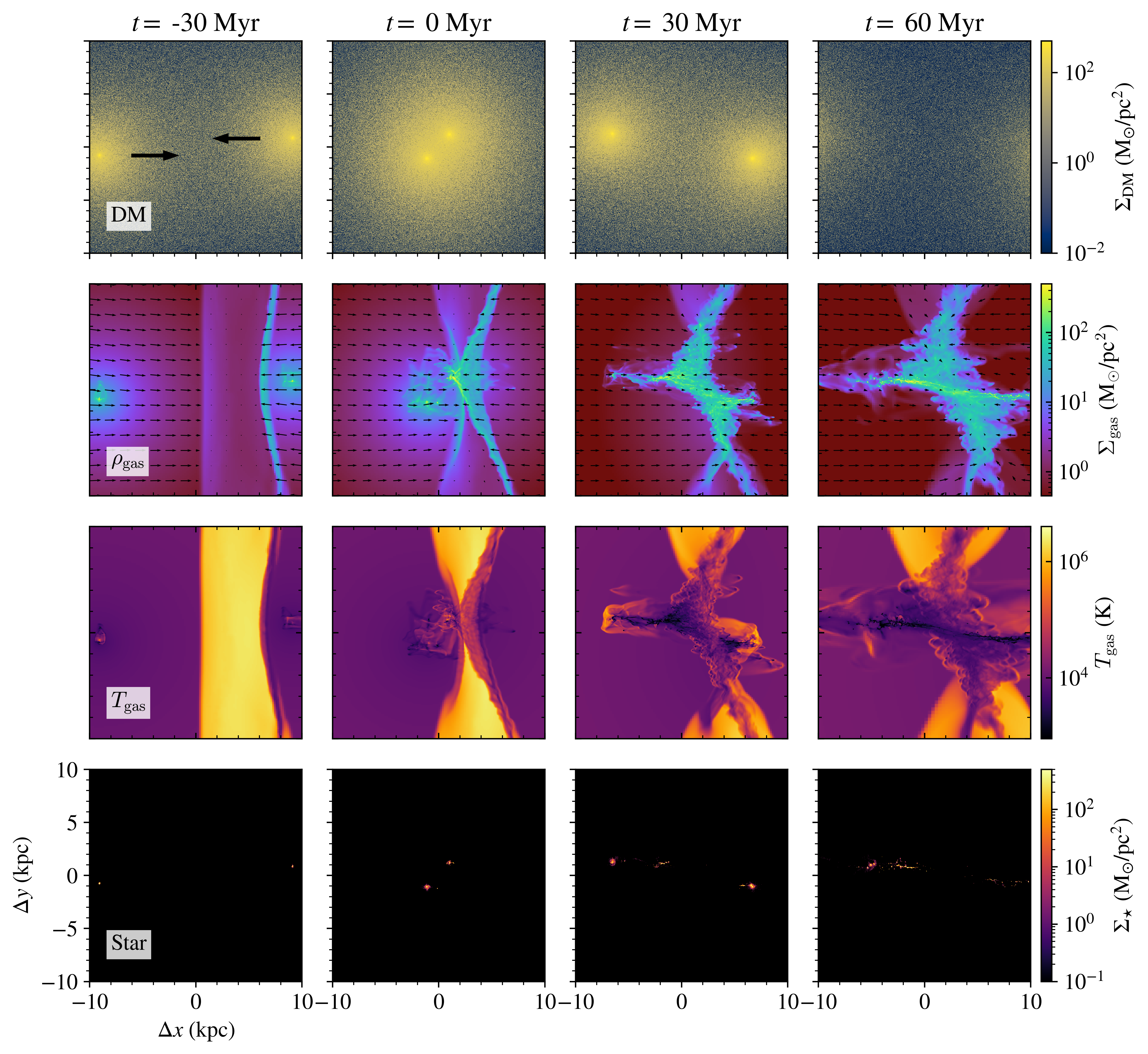}
    \caption{
        \emph{\textbf{Illustrative Mini-Bullet Simulation: collision in-progress (collision-plane view).}}
        Same as Figure \ref{fig:1} but zoomed-in in time and space to show snapshots of the galaxies in the midst of their collision, $t = -30,\;0,\;30,\;60$ Myr.
        The surface density of dark matter (first row), surface density of gas with velocity vectors overplotted (second row), density-weighted average gas temperature (third row), and surface density stars -- just those formed after the start of the simulation -- (fourth row) are presented.
        Gas velocity vectors are projected onto the image plane, which is the $xy$ plane to which the centers of mass of the colliding-galaxy orbits are confined.
        The velocity vectors are proportional to the length of the velocity vector arrows and the largest arrow corresponds to $\sim$300 $\kms$.
        Black arrows in the top-left panel show the progenitors' moving directions.
        Dark matter and stars are projected in 50 kpc-depth layers and gas is projected in 10 kpc-depth layers.
        \label{fig:2}
        }
\end{figure*}

\begin{figure*}[t]
    \centering
    \includegraphics[width=0.95\textwidth]{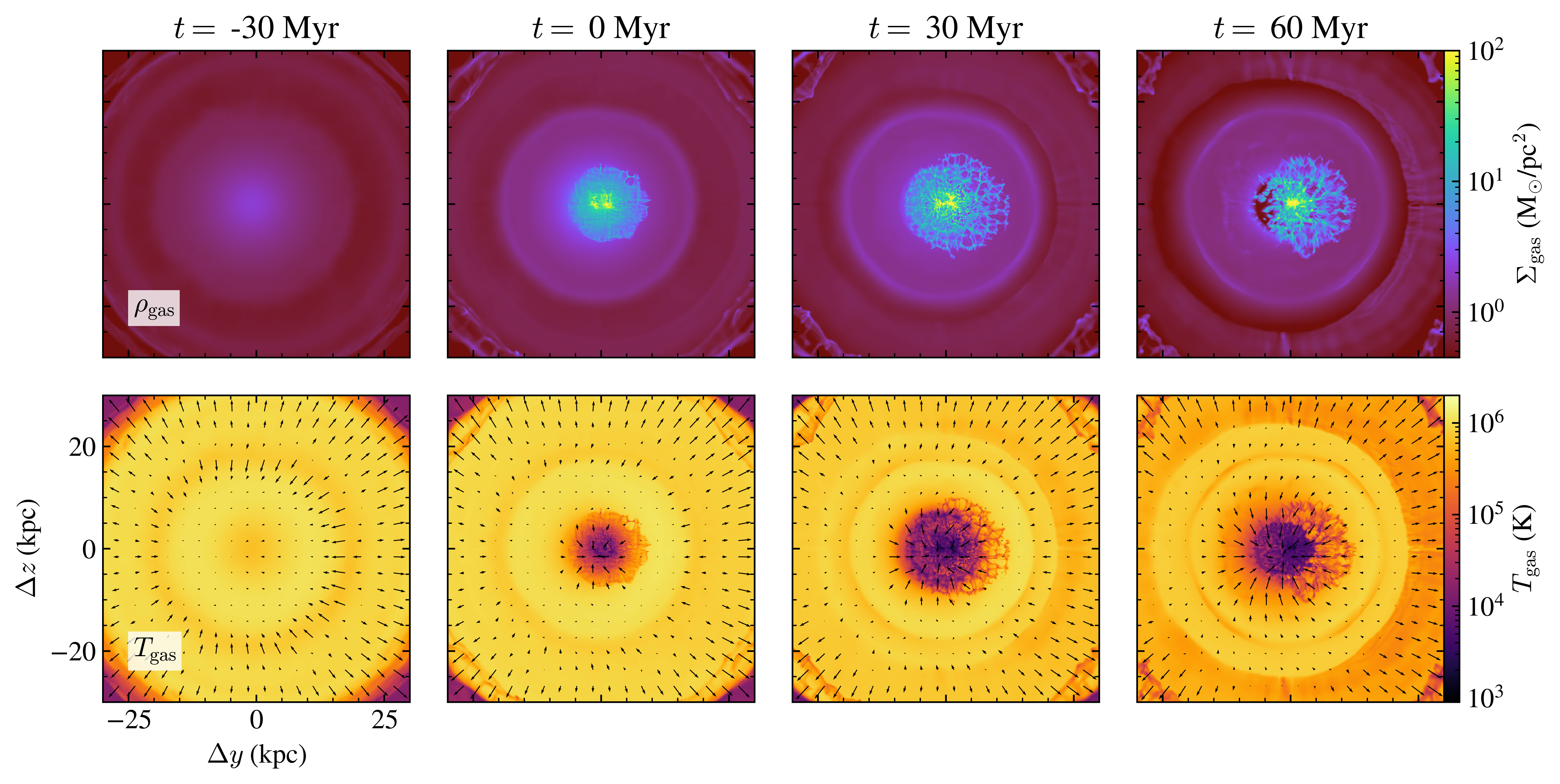}
    \caption{
        \textbf{\emph{Illustrative Mini-Bullet Simulation: collision in-progress (viewed perpendicular to collision-plane).}}
        Same simulation snapshots as Figure \ref{fig:2}, at $t = -30,\;0,\;30,\;60$ Myr, but viewed in projection along the $x$-axis in the $yz$-plane -- i.e. the view close to the line of collision.
        The surface density of the gas (top row) and density-weighted average gas temperature with velocity vectors overplotted (bottom row) are presented.
        Gas velocity vectors are projected onto the image plane, which is the $yz$ plane, in which the center of the mass of the colliding progenitor system was at $t=0$ Myr.
        The velocity vectors are proportional to the length of the velocity vector arrows and the largest arrow in each panel corresponds to $\sim$200 $\kms$.
        Gas is projected in 10 kpc-depth layers.
        \label{fig:3}
        }
\end{figure*}

In the idealized galaxy-galaxy collision simulations, we use the publicly available adaptive mesh refinement (AMR) code {\sc enzo} \citep{Bryan14, Brummel-Smith19} with its hydrodynamics solver {\sc zeus} \citep{Stone92a, Stone92b}.
The {\sc grackle} library \citep{Smith17} is used to compute radiative gas cooling and heating assuming the homogeneous ultraviolet (UV) background of \cite{HM12} at $z=0$, by interpolation of the lookup table generated from the {\sc cloudy} code \citep{Cloudy17}\footnote{We tested with simulations and verified that adopting the UV background at $z=1$, instead, does not alter the gas cooling and heating and star formation physics significantly.}.
We refine the simulation box of $(2.621 \; {\rm Mpc})^{3}$ down to a spatial resolution of $\Delta x = 10$ pc (level $l_{\rm max} = 12$), which is eight times coarser than the simulations carried out in \citetalias{Lee21}.  
It is necessary here to reduce the computational cost of simulating a larger box over a longer time than before, so we must relax the resolution, aware that it cannot resolve star cluster formation as was a primary goal of \citetalias{Lee21}, with its spatial resolution of $\Delta x = 1.25$ pc, but is sufficient to resolve the galaxy formation properties we need here.
Our refinement strategy is super-Lagrangian, meaning if a cell contains more mass than the mass threshold (i.e. its mass density exceeds a density threshold), that cell splits into eight child cells.
At refinement level $l$, for gas with star formation threshold gas number density $n_{\rm th}$, 
$M^{l}_{\rm ref,\,gas} = 2^{-0.333 (l-12)}\times M^{12}_{\rm ref,\,gas}$, where $M^{12}_{\rm ref,\, gas} = 10000\; {\msun} = n_{\rm th} \Delta x^{3} \simeq 2.5 \; M_{\rm Jeans}(T=100\,{\rm K}, \; n=400\,{\rm cm}^{-3})$, 
and for (dark matter and stellar) particles, $M^{l}_{\rm ref,\,part} = 2^{-0.107 (l-12)}\times M^{12}_{\rm ref,\,part}$, where $M^{12}_{\rm ref,\,part} = 16000 \; {\msun}$.
We note that, while refinement proceeds down to a cell size that is small enough not to contain more than a fixed physical mass of either gas or particles, this refines the force length resolution of the gravity and gas pressure forces but does not refine the particle mass resolution.  Dark matter and stellar particles (after releasing stellar feedback) have a fixed mass, at all times and do not refine.

To model feedback-regulated star formation, we adopt subgrid algorithms for under-resolved small-scale physical processes, just as we did in \citetalias{Lee21}, but with some adjustments for coarser resolution and an assumption of higher thermal energy released by SNe associated with the massive stars.  A parcel of gas is determined to form stars, according to the approach in \citet{Cen92}, and the outcome of that star formation is assumed to follow the star-forming molecular cloud model \citep[SFMC; for details, see][]{Kim13, Kim19}. 
In brief, an SFMC particle is created when the following criteria are met:
(1) the density of a gas cell exceeds $n_{\rm th} = 400 \; {\rm cm}^{-3}$, 
(2) the gas flow is converging,
(3) the cooling time of the cell is shorter than its dynamical time ($t_{\rm dyn}$), 
and (4) the mass in that cell is enough to create an SFMC particle heavier than $m_{\rm SFMC} = 5 \times 10^{3} \msun$ (which leads to a ``permanent'' star particle mass -- i.e. total mass that follows the initial mass function of the stars that form, of $m_{\star,\, \rm new} = 10^{3} \msun$).
The SFMC particle returns 80\% of its original mass to the gas in a time equal to $12 t_{\rm dyn}$, according to our assumed star formation efficiency for converting gas-to-star mass \citep{Krumholz07}, along with supernova thermal feedback of $10^{51}$ erg per $50 \msun$ of permanent star mass that peaks at $1t_{\rm dyn}$ and 2\% of metal yield \citep[see also][]{Kim11}.

We follow the method used in \citetalias{Shin20} and \citetalias{Lee21} to initialize two progenitor galaxies by utilizing the {\sc dice} code \citep{Perret16}.
We place two identical progenitor galaxies 60 kpc apart and set their relative velocity to be $300 - 600$ $\kms$.
Since the observed line-of-sight velocity difference of DF2 and DF4 is 358 $\kms$ \citep{vD22}, relative collision velocities need to be higher than that.
The pericentric distances $r_{\rm min}$ is set to be 2 kpc.

While the parameters for these idealized galaxy-galaxy collisions in Step 1 are similar to those adopted in
\citetalias{Shin20}, there are several important distinctions between the simulations here in Step 1 and those in \citetalias{Shin20}.   
First, the progenitors here are taken to be spherical halos with gaseous baryons, which self-consistently form stars before their collision,
while in \citetalias{Shin20}, the intention was to model the collision of present-day galaxies with well-established disks.
Second, the spatial resolution here is much higher than in \citetalias{Shin20}, which makes a difference in the ability of
the collision simulations to produce the post-collision stellar systems.  To model the NGC1052 galaxy group, it is
necessary to demonstrate that a single collision with realistic parameters can naturally produce a trail of 
$\sim$10 DMDGs.
The coarser spatial resolution of simulations in \citetalias{Shin20}, of 80 pc 
(as opposed to this paper's 5 pc resolution) prevented us from forming $\sim 10$ DMDGs there, however. 
In those lower-resolution simulations, fewer objects of higher mass were formed, in general, and it was necessary
to tune the choice of parameters just so as to maximize this number.  Even so, only up to 6 DMDGs were formed.
In the current paper, we believe we have converged with spatial resolution, as demonstrated by the comparison of the
runs with 5 and 10 pc resolution, respectively.  
And the formation of $\sim 10$ DMDGs did not require such 
fine-tuning as before, either.

\citetalias{Lee21}, on the other hand, had even higher spatial resolution than the simulations here, but only by
applying the AMR to a limited ``zoom-in'' region surrounding the most
massive DMDG formed by an idealized galaxy-galaxy collision.   
As such, it did not address the questions
at issue here, of producing a trail of $\sim$10 DMDGs.

The progenitors are initialized with only gas and dark matter. Their dark matter halos have $M_{200}=1.66\times10^{10} \msun$, $J_{200}=1.03\times10^{12} \msun\;{\rm kpc}\;\kms$, and $R_{200}=51.7$ kpc, and follow the NFW profiles \citep{NFW}, with concentration parameter $c=13$\footnote{
The value $c=13$ was chosen to match the value adopted for the simulations in \citetalias{Shin20}.}.
For the gas density profiles, instead of initializing the progenitor galaxies with exponential disks as in \citetalias{Shin20} and \citetalias{Lee21}, this time, we adopt the pseudo-isothermal (PIS) profile without any rotation,
\begin{equation}
    \rho(R) = \rho_{0} \frac{1}{1 + (R/R_{\rm s})^{2}},
\end{equation}
with a scale radius of $R_{\rm s,\, gas} = 2$ kpc, which is more commonly observed and used to simulate low-mass satellite galaxies, to initialize the gas density profile of the progenitors \citep[e.g.,][]{Kurapati20}. 
The total mass is $1.89\times10^{10} \msun$, and the gas fraction is $f_{\rm gas} = M_{\rm gas} / M_{\rm total} = 12.4\%$, with no stars at the beginning.\footnote{While this gas fraction is between the cosmic mean baryon fraction \citep{Planck20} and the typical baryon fraction in low-mass galaxies, it lies within the observed scatter for low-mass galaxies \citep[cf.][]{Wechsler18, Crain23}. 
Our choice is motivated by the original suggestion of the {\it Mini-bullet} model \citep{Silk19} which involved gas-rich colliding progenitors.}
The gas is initially set to a temperature of $10^{4}$ K and metallicity of $Z = 0.1 \zsun = 0.002041$ to match the metallicity of a low-mass galaxy at $z \sim 1 - 2$.
We use $2 \times 10^{6}$ dark matter particles to model each progenitor, resulting in mass resolution of $m_{\rm DM} = 8.29 \times 10^{3} \msun$.

We perform a suite of four galaxy-galaxy collision simulations, each with a different set of initial configurations. 
In Table \ref{tab:1}, we summarize this set of structural parameters we test and the stellar masses of the most massive DMDG that result, the velocity difference of the most and second massive DMDGs (corresponds to DF2 and DF4), and the number of DMDGs produced at $t_{\rm end}$, after the collisions.
Among the tested results, it is notable that there are collision configurations that result in a large velocity difference ($\sim 300$ $\kms$) between the first and second most massive DMDGs that form which is similar to that between DF2 and DF4 \citep[][]{vD22}. 
We note that, in these idealized collision simulations, the post-collision separation velocity does not increase with time, while in the more realistic simulations we will describe below, in which the progenitors are also satellites of the massive galaxy NGC1052, their post-collision separation velocity can increase with time.   
As a result, it is possible that their immediate post-collision separation velocity was somewhat less than observed now for DF2 and DF4.

The time history of the mini-Bullet system from pre- to post-collision is illustrated by Figures \ref{fig:1}, \ref{fig:2}, and \ref{fig:3}, which show snapshots of the results of our {\sc enzo} simulation for Run 3 in Table \ref{tab:1}. 
At $t=-110$ Myr, two progenitor galaxies are initialized with a separation of 60 kpc, approaching at 500 $\kms$.
Figure \ref{fig:1} zooms out for time-slices long before, during, and long after the progenitor collision, to show how
the {\it Mini-bullet} collision dissociates collisionless components, dark matter and stars, of these progenitors from their gas ($t=90$ Myr in Figure \ref{fig:1}).  
In Figure \ref{fig:2}, we zoom in, both in time and space, to show the immediate effects of the collision on the two progenitors and the distinctions between those effects on their collisionless components (dark matter and stars), which can pass through each other, versus their collisional component, their gaseous baryonic interstellar media (ISM), which, as a fluid, cannot.  
Instead, the ISM of each progenitor is halted from its supersonic approach to its counterpart in the other progenitor, by a strong shock that heats the gas to $T > 10^{7}$ K, in which the sound speed is comparable to the collision velocity ($t=-30$ and 0 Myr in Figures \ref{fig:2} and \ref{fig:3}).
There are two shocks, one on each side of their collision center, one to decelerate each incoming progenitor's ISM, producing a layer of shock-heated gas between them, whose mass grows as the collision proceeds to overtake more and more of each progenitor's ISM. 
The shock-heated gas begins cooling radiatively as soon as it is heated, and the shock becomes a radiative one, for which the cooling time is short compared with the time for the shock velocity to change ($t=30$ Myr in Figures \ref{fig:2} and \ref{fig:3}).  
In such shocks, the gas can cool isobarically, to well below $10^4$ K, increasing its density in inverse proportion to the temperature, so its density increases as the square of the shock Mach number.
Since the gas in the ISM of each progenitor is centrally-concentrated, the cooling time is shortest for shocked gas that was initially closer to the centers of the progenitors, so this is where the shocks become radiative first, and the cooling gas is subject to dynamical and gravitational
instabilities. 
For this off-axis collision, dense gas clumps are formed due to radiative cooling near the line that connects the progenitors ($y=0$ line). 
This results in a combination of cold gas which collapses towards their centers while surrounding shells of still-hot shocked gas are driven to expand away from the line of collision, by pressure forces, as shown in Figure \ref{fig:3}, from the view along the line of collision ($x$-axis).  
After star formation in the collision-induced dense gas clumps, we identify seven DMDGs that are roughly aligned between the progenitors, forming an ``S-shaped'' trail ($t=390$ Myr in Figure \ref{fig:1}).
Gas as the fuel of star formation is almost exhausted before this time, preventing star formation in the DMDGs on the trail.
The trail of DMDGs is tidally stretched over time by the gravitational field of the separating progenitors at each end, which retain their dark matter and stars, making the ``S-shaped'' trail longer as the separation between galaxies grows ($t=690$ Myr in Figure \ref{fig:1}).

\subsection{Producing ``Tailor-Made'' initial conditions---backward orbit integration} \label{sec:2.2}

Equipped with this knowledge from our suite of idealized galaxy-galaxy collision simulations regarding which parameters are suitable for producing DMDGs like DF2 and DF4, we now move to establish a realization of colliding progenitor pairs which are satellites of a massive host galaxy.
This realization should produce two DMDGs, each with $M_{\star} \gtrsim 10^{8} \msun$ (for DF2 and DF4) and several UDGs along a line after $\sim 8$ Gyr, to match the observations \citep{Roman21, vD22}.  
These observations will include the radial separations and positions on the sky of DF2 and DF4, their radial velocity separation, and their radial velocities relative to that of NGC1052.

The unknown initial conditions that will lead to these final conditions for DF2 and DF4 are the pre-collision locations and velocity vectors of the collision progenitors, relative to each other and to NGC1052.  
To derive these we will integrate the orbits of DF2 and DF4 backward in time to locate the collision that produced them and their velocity vectors at that time.   
By doing a large enough sample of such orbit integrations, we were able to identify those for which the orbits of DF2 and DF4 once crossed in the past, indicating the time and place of the collision we postulate to have formed them. 
From this subset of the orbit integrations that led to collisions in the past, we refined our sample further, to those for which the lookback time of the collision was large enough to explain the observationally inferred ages of the GCs of DF2 and DF4, i.e. $\sim 8$ Gyr \citep{J.Ma20, Fensch19, vD18c}.

We noticed that, for this refined set of cases, the relative velocities of DF2 and DF4 immediately after the collision that produced them, were typically lower than the presently-observed radial separation velocities, by $\gtrsim 100$ $\kms$, reflecting the relative acceleration of their post-collision orbits over time.
This means that the immediate post-collision separation velocities were consistent with the collisions in Runs 2 and 3 in Table \ref{tab:1}, thereby confirming that such post-collision outcomes were, indeed, possible from suitably high-velocity galaxy-galaxy collisions.

The next step was to select three of these orbit integrations that yielded the collisions at the right lookback time, and determine the pre-collision parameters of the progenitor galaxies that would produce these immediate post-collision separations and velocities for DF2 and DF4.  
In principle, if we had a large enough sample of cases in Table \ref{tab:1}, we might have had a close enough match to the orbit integrations that we could use them to identify the pre-collision configuration that would lead to the post-collision configuration derived from a given orbit integration.
In practice, however, it is computationally prohibitive to run so many cases of idealized collisions for this purpose.   
Instead, we determined the pre-collision configuration as follows.

The exact pre-collision configurations of the two progenitors are decided by setting relative collision velocities $v_{\rm col}$ to two times the relative velocities of DF2 and DF4 $\Delta v_{\rm DF2-DF4}$, resulting in the two progenitors collide with $\sim 500$ $\kms$, and pericentric distances of $r_{\rm min} \sim 2$ kpc.
This follows the observation from the simulations presented in Section \ref{sec:2.1} that the velocity difference of the first and second most massive DMDGs formed in  $\Delta v_{\rm DMDG,\,max2}$ is roughly a half of $v_{\rm col}$ of the progenitors $\Delta v_{\rm DF2-DF4} = \Delta v_{\rm DMDG,\,max2} \sim 0.5 v_{\rm col}$
(with ignorance of the long-term velocity change due to the presence of the host gravity; refer to Section \ref{sec:2.1} and Table \ref{tab:1}).

Finally, with the immediate pre-collision configurations derived in this way for each of the three selected cases above, another backward time integration was required for each, this time of the orbits of their two progenitor galaxies, destined to meet at their collision moment, to place them at large separations at earlier times, from which we could perform full gravitohydrodynamic simulations of their collision events, from start to finish.
Now we will describe these steps in more detail.

We use the Integrator with Adaptive Step-size control, 15th order (IAS15), a gravitational dynamics integrator, implemented in the {\tt Rebound} orbit integration code \citep{Rein12, Rein15} to backtrace the orbits of DF2 and DF4 based on the currently observed quantities and find the collision point of the progenitor galaxies, where DF2 and DF4 should have started to form at the same time.
Then, using the confined collision parameters that can produce multiple aligned DMDGs (Section \ref{sec:2.1}), we place progenitor galaxies to make initial conditions for the hydrodynamic simulation (Section \ref{sec:2.3}).
Finding the collision point is not trivial, because depending on the initial locations and motions, the back-traced DF2 and DF4 may not collide.
Therefore, we should find the initial conditions from the parameter space that consists of the range of current locations and motions of DF2 and DF4\footnote{ We note that this assumes that the three-body system of the two progenitors and the host galaxy NGC1052
can be treated as an isolated system for a cosmologically long time interval.
This is a good approximation, as long as the mass of the group, which is dominated by NGC1052, did not substantially
evolve over that time.   Since we start at $z\sim1$, a typical mass assembly history for an object like 
NGC1052 would only have increased its mass by $\sim50\%$ by the present, so this is a reasonable approximation.}.

\begin{figure}[t]
    \centering
    \vspace{0mm}  
    \includegraphics[width=0.47\textwidth]{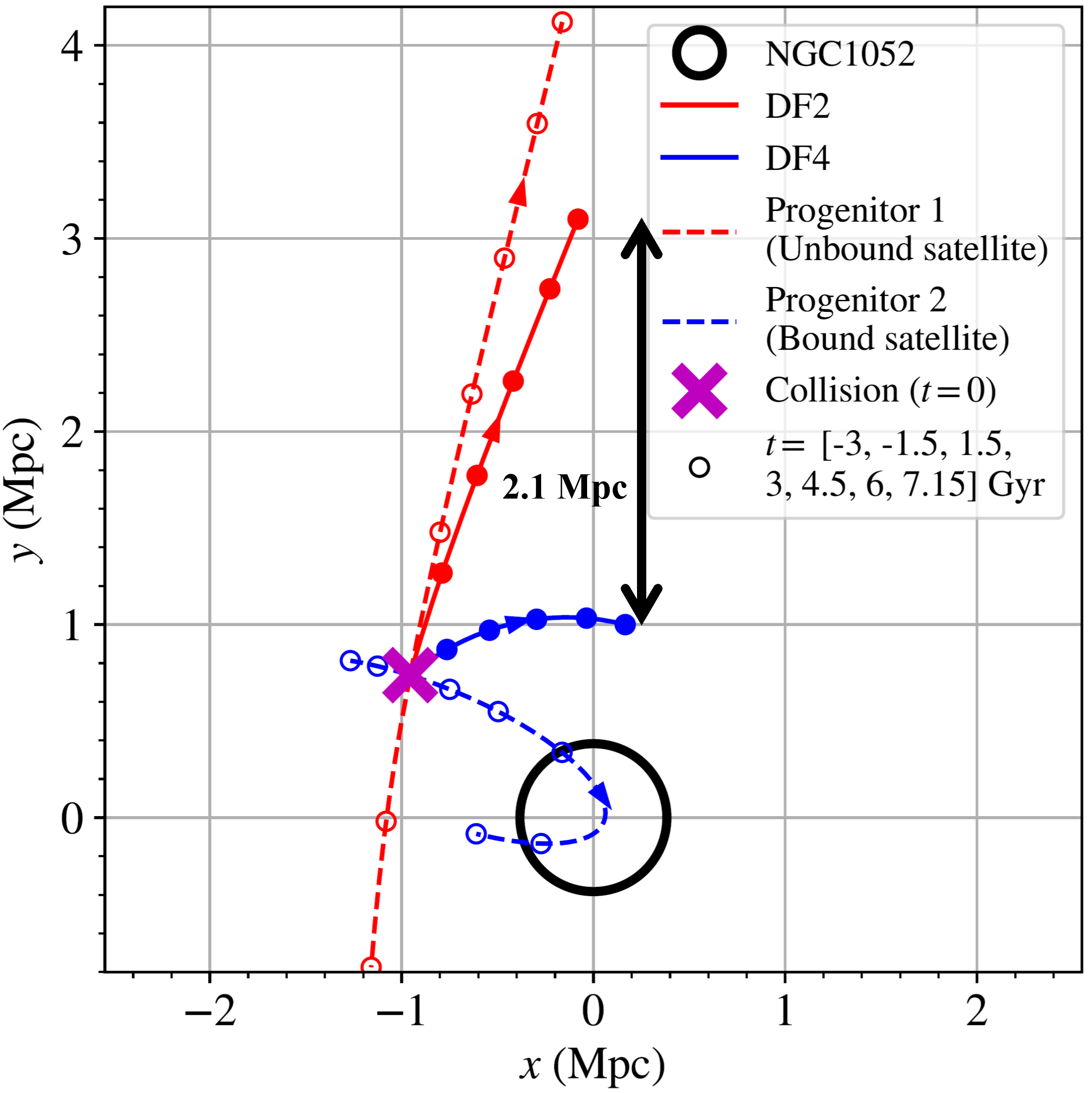}
    \caption{\label{fig:4}
        \emph{\textbf{Backward Orbit Integration to Produce a ``Tailor-Made'' Initial Condition.}}
        An example of backtraced orbits of DF2 (solid red) and DF4 (solid blue), computed orbits of the progenitors (dashed red and blue) that are expected to produce a trail of DMDGs including DF2 and DF4 in a {\textit {Mini-bullet}} collision, and their expected orbits after the collision.
        The purple cross indicates the location of the {\textit {Mini-bullet}} collision.
        The black circle is the virial radius ($R_{200} = 400$ kpc) of the host, NGC1052.
        This example corresponds to {\tt TM3} run in Table \ref{tab:2}
        See Section \ref{sec:2.2} for more information.
    }
    \vspace{0mm}
\end{figure}

In Figure \ref{fig:4}, we present an example of the three initial conditions based on the orbit integration calculation results.
We place a dark matter-only massive host (black circle) with a halo mass of $M_{\rm DM} = 6.12 \times 10^{12} \msun$ to represent the host, NGC1052, at the center \citep{Forbes19}.
The orbits of DF2 and DF4 with the stellar mass of $M_{\star} = 2 \times 10^{8} \msun$ need to be time-reversed integrated to find the collision point.
The radial distances of DF2 and DF4 are constrained with a certain amount of uncertainties ($22.1\pm1.2$ Mpc for DF2 and $20.0\pm1.6$ Mpc for DF4) \citep{vD18a, vD19, vD22}\footnote{However, there have been debates about the robustness of the distance measure method \citep{Trujillo19, Monelli19}}, but the distance to NGC1052 is not well measured.
Thus, in this study, we will assume the distance to NGC1052 as 19 Mpc consistent with previous studies using the surface brightness fluctuation method \citep{Tonry01, Blakeslee01} and the Virgo-infall corrected radial velocity \citep{GildePaz07}\footnote{Note that these previous distance measurements yield $d=19-21$ Mpc and we only test $d=19$ Mpc case in this study. Thus, the distance to NGC1052 needs to be measured precisely in future observation, which in turn will require follow-up simulation adopting the observed distance.}.
The transverse distances of DF2 and DF4 from NGC1052 can be simply calculated from their angular separations \citep{Roman21}: $-0.081$ Mpc (DF2) and $+0.168$ Mpc (DF4) assuming the observed distance to DF2 and DF4.
For the radial distances, we set errors to be 0.25$\times$(observation error) and for the transverse distances, we adopt 0.002 Mpc as errors.
The three galaxies and the observer are not exactly on the same plane but we ignore the deviation and assume that the transverse (proper) and radial (line-of-sight) distances can be used as the $x$-axis and the $y$-axis in our orbit integration as if they are on the same plane.
The radial velocities of DF2 and DF4 relative to NGC1052 are also measured with unspecified errors \citep{vD22}: $+315$ $\kms$ (DF2) and $-43$ $\kms$ (DF4).
Since the observation errors for radial velocities are not specified, we set their errors to be 10 $\kms$.
For those known parameters (radial and transverse distances and radial velocities), we let them vary within observed values $\pm$ errors.
However, since the transverse motion of DF2 and DF4 cannot be observed, we vary the transverse velocities within fixed ranges: 
[10 $\kms$, 170 $\kms$] for the progenitor 1 (bound satellite, blue dashed in Figure \ref{fig:4}) and [0 $\kms$, 80 $\kms$] for the progenitor 2 (unbound satellite, red dashed in Figure \ref{fig:4}) with an interval of $\Delta v = 2.5\;\kms$.
To confine the parameter space, we uniformly sample the parameter space, testing $> 1,500,000$ cases, and find $\sim 3000$ cases where DF2 and DF4 started from places closer than 10 kpc, the place of a {\it {Mini-bullet}} galaxy collision.

After finding the collision point, we place two progenitor galaxies (blue and red circles on dashed lines) with dark matter halo mass of $M_{\rm DM} = 1.42 \times 10^{10} \msun$ and $M_{\rm gas} = 2.47 \times 10^{9} \msun$ ($f_{\rm gas} = 0.148$) with no stars\footnote{This is not realistic but in orbit integration calculation, we treat them as point masses, thus making no difference.}, which is similar to the mass we tested, using information from the idealized two-galaxy collision simulations in Section \ref{sec:2.1}.
One of the progenitors (dashed blue) is gravitationally bound to the host and has a highly eccentric orbit.
The other progenitor (dashed red) is not bound to the host but comes from outside of the NGC1052 system with a high velocity of $> 400$ $\kms$ relative to the host, which is higher than the host virial velocity $V_{200} = 260\;\kms$.

\begin{deluxetable*}{cccccccccc}[t]
    \tablenum{2}
    \vspace{-1mm}
    \tablecaption{
    Initial conditions for simulations of satellite-satellite galaxy collisions around a massive host with $M_{\star} = 6.12 \times 10^{12} \msun$.
    The initial conditions are ``tailor-made'' in attempt to produce collisions whose outcome matches observations of the NGC1052 group \citep{vD22}, including backward orbit integration described in Section \ref{sec:2.2}.
    \label{tab:2}
    }
    \vspace{-1mm}
    \tablewidth{0pt}
    \tablehead{
        \colhead{Run name} & \colhead{$v_{\rm col}$} & \colhead{$r_{\rm min}$} & \colhead{$R_{\rm s,\,gas}$} & \colhead{$c_{\rm halo}$} & \colhead{($x_0$, $y_0$)$_{\rm DF2}$} & \colhead{($v_{x0}$, $v_{y0}$)$_{\rm DF2}$} & \colhead{($x_0$, $y_0$)$_{\rm DF4}$} & \colhead{($v_{x0}$, $v_{y0}$)$_{\rm DF4}$} & \colhead{Orbit case}
    \\[-2mm]
        \colhead{{\tt TM}} & \colhead{($\kms$)} & \colhead{(kpc)} & \colhead{(kpc)} & & \colhead{(Mpc)} & \colhead{($\kms$)} & \colhead{(Mpc)} & \colhead{($\kms$)}
    \\[-2mm]
        \colhead{({\tt ``Tailor-Made''})} & & & & & & & &
    \\[-2mm]
        \colhead{(1)} & \colhead{(2)} & \colhead{(3)} & \colhead{(4)} & \colhead{(5)} & \colhead{(6)} & \colhead{(7)} & \colhead{(8)} & \colhead{(9)} & \colhead{(10)}}
    \startdata
    {\tt TM1} & 531 & 2 & 3 & 7 & (-0.079, 3.1) & (105, 325) & (0.167, 1.0) & (150, -33) & 1\\
    {\tt TM2} & 468 & 2 & 3 & 7 & (-0.081, 3.1) & (95, 305) & (0.169, 1.0) & (140, -33) & 2 \\
    {\tt TM3} & 529 & 2 & 3 & 7 & (-0.080, 3.1) & (125, 305) & (0.167, 1.0) & (170, -43) & 3\\
    {\tt TM4} & 531 & 1.8 & 3 & 13 & (-0.079, 3.1) & (105, 325) & (0.167, 1.0) & (150, -33) & 1\\
    {\tt TM5} & 531 & 1.8 & 3 & 7 & (-0.079, 3.1) & (105, 325) & (0.167, 1.0) & (150, -33) & 1\\
    {\tt TM6} & 531 & 1.5 & 4 & 13 & (-0.079, 3.1) & (105, 325) & (0.167, 1.0) & (150, -33) & 1\\
    {\tt TM7} & 531 & 1.8 & 4 & 13 & (-0.079, 3.1) & (105, 325) & (0.167, 1.0) & (150, -33) & 1\\
    {\tt TM8} & 531 & 2 & 4 & 13 & (-0.079, 3.1) & (105, 325) & (0.167, 1.0) & (150, -33) & 1\\
    \enddata   
    \tablecomments{
        (1) run name,
        (2) relative velocity of the two progenitors at pericenter,
        (3) pericentric distance of the progenitors, 
        (4) gas scale radius in the PIS profile,  
        (5) concentration parameter of progenitor dark matter halos,
        (6) transverse ($x$) and radial ($y$) coordinates relative of DF2 to NGC1052,
        (observed radial distance from the earth is $20.0$ Mpc),
        (7) transverse velocity ($v_x$) and radial velocity ($v_y$) of DF2 relative to NGC1052 (presently observed to recede from NGC1052 at 315 $\kms$; \citealp{vD22}),
        (8) transverse ($x$) and radial ($y$) coordinates of DF4 relative to NGC1052, (observed radial distance from the earth is $22.1$ Mpc),
        (9) transverse velocity ($v_x$) and radial velocity ($v_y$) of DF4 relative to NGC1052 (presently observed to approach NGC1052 at $-43$ $\kms$; \citealp{vD22}),
        (10) orbits of DF2 and DF4 in backward orbit integration calculation described in Section \ref{sec:2.2}.
        The difference in the orbits originates from different starting positions and velocities of DF2 and DF4 in the time-reversed orbit integrations, as listed in columns (6)-(9), accounting for observational error and uncertainties in transverse velocities of DF2 and DF4, which cannot be observed.
        Runs ({\tt TM4$-$8}) share the starting positions and velocities of DF2 and DF4 that define Orbit 1 with {\tt TM1}. To be clear, all these ``starting'' values for backward time integration are actually ``final'' values when time is expressed as going forward to the present epoch.}
    \vspace{-8mm}
\end{deluxetable*}

Using these orbit integration results, we set up initial conditions consisting of the massive host and two progenitors at 0.3 Gyr before their {\it {Mini-bullet}} collision,  to make them relax after the artificial starbursts that occur after initialization.
Their gas density distributions are given by the PIS profile, while their dark matter densities follow an NFW profile.
Using the {\sc dice} code \citep{Perret16}, the density profile of NGC1052-like host is sampled with $10^{6}$ dark matter particles ($m_{\rm DM,\, host} = 6.1 \times 10^{6} \msun$) and the colliding progenitors' NFW profile is realized with $10^{7}$ dark matter particles ($m_{\rm DM,\,prog} = 1.42 \times 10^{3} \msun$) each.
Since the lookback time of 8 Gyr corresponds to $z \sim 1$, we will now set the progenitors' halo concentration to be $c = 7$, about a factor of two lower than $c = 13$, the value which was adopted in the idealized simulations in Table \ref{tab:1}. 
This is because, for a halo with a given mass, the concentration parameter is lower for a halo that formed earlier.
We summarize sets of collision configurations and structural parameters of the progenitors we test in Table \ref{tab:2}.  
Henceforth, we shall refer to all of the initial conditions described here in Section \ref{sec:2.3} as ``Tailor-Made'' and the cases in Table \ref{tab:2} as ``Tailor-Made'' cases
(e.g. {\texttt {\textit {TM1}}} shall refer to \emph{Tailor-Made case 1}).
Among those initial conditions, we specifically focus on {\tt TM1}, {\tt TM2}, and {\tt TM3} for our analysis presented in Section \ref{sec:3.1} and \ref{sec:3.2},
as representative runs that start from different pre-collision orbital parameters of the progenitor galaxies and yield results that reproduce the observations.  
(Note: {\tt TM5} is a variation on {\tt TM1}, with different separation distance at
the closest approach for the colliding progenitors, but with a similar enough outcome
to {\tt TM1}'s that we do not give a more detailed description here.

\subsection{Mini-bullet satellite-satellite galaxy collision and the
formation of multiple dark matter-deficient galaxies} \label{sec:2.3}

With these Tailor-Made initial conditions, summarized in Table 2, we again use the {\sc enzo} code to simulate the high-velocity collision of two satellite galaxies and the orbital evolution of the progenitors and resultant DMDGs after the collision.
Hydrodynamics and star formation physics are the same as the idealized simulations in Section \ref{sec:2.1}.
The box size is increased to (5.243 Mpc)$^{3}$ to fully capture the orbits of the product DMDGs, but the nested refinement region is set to resolve only the progenitors and product DMDGs, not the massive host galaxy.
Static refinement up to level 3 ($\Delta x = 10.24$ kpc) is applied outside the nested refinement regions.
We change the refinement scheme due to the refinement level change ($\Delta x = 5$ pc at level $l_{\rm max} = 14$) and refine more for the particles to resolve star-dominated structures better: 
at refinement level $l$, for gas, 
$M^{l}_{\rm ref,\,gas}=2^{-0.264 (l-14)}\times M^{14}_{\rm ref,\,gas}$, where $M^{14}_{\rm ref,\, gas}=4000\; {\msun} \simeq 2.5 \; M_{\rm Jeans}^{100 \,{\rm K}}$, 
and for particles, $M^{l}_{\rm ref,\,part}=2^{-0.273 (l-14)}\times M^{14}_{\rm ref,\,part}$, where $M^{14}_{\rm ref,\,part} = 4000 \; {\msun}$.
Following the change of cell spatial resolution and refinement criteria, we change the star formation density threshold to $n_{\rm th} = 1.6 \times10^{3} \; {\rm cm}^{-3}$ and the SFMC particle mass threshold to $m_{\rm SFMC} = 2.5 \times 10^{3} \msun$ (permanent star particle mass of $m_{\star,\, \rm new} = 5 \times 10^{2} \msun$).
The simulations are run for 2.3 Gyr, 0.3 Gyr before the collision, and 2 Gyr after the collision at which we analyze the properties of the product DMDGs.
Due to the significant computational time needed to run the full $\sim 8$ Gyr of simulation, we further supplement the simulations with orbit integration code {\tt Rebound} for the remaining $\sim 6$ Gyr to track the orbital evolution of the DMDGs and colliding progenitors during later stages.

\section{Results} \label{sec:3}

In this section, we present the results of the idealized satellite-satellite galaxy collision simulation introduced in Section \ref{sec:2.3} and discuss how the product DMDGs can be compared to the observation and what the physical properties of the DMDGs are.

\subsection{Orbits of the produced galaxies: a trail of dark matter-deficient galaxies and progenitors} \label{sec:3.1}

\begin{deluxetable*}{cccccccc}[t]
    \tablenum{3}
    \vspace{-1mm}
    \tablecaption{
    Results from a suite of simulations of satellite-satellite galaxy collisions around a massive host.
    See Section \ref{sec:3.1} for a description and discussion of the properties of the product DMDGs.
    \label{tab:3}
    }
    \vspace{-1mm}
    \tablewidth{0pt}
    \tablehead{
        \colhead{Run name} & \colhead{$M_{\rm \star, \, DMDG,\,max2}$} & \colhead{$\Delta v_{\rm DMDG,\,max2}$} & \colhead{$\Delta v_{\rm DMDG,\,max2,\,now}$} & \colhead{$N_{\rm DMDG}$} & \colhead{$t_{\rm end,\,sim}$} & \colhead{$t_{\rm end,\,orbit}$} & \colhead{Matches}
    \\[-2mm]
        \colhead{({\tt TM(``Tailor-Made'')})} & \colhead{($10^8\msun$)} & \colhead{($\kms$)} & \colhead{($\kms$)} & & \colhead{(Gyr)} & \colhead{(Gyr)} & \colhead{NGC1052 group?}
    \\[-2mm]
        \colhead{(1)} & \colhead{(2)} & \colhead{(3)} & \colhead{(4)} & \colhead{(5)} & \colhead{(6)} & \colhead{(7)} & \colhead{(8)}}
    \startdata
    {\tt TM1} & 2.6, 1.5 & 216 & 324 & 11 & 2 & 8.375 & Yes\\
    {\tt TM2} & 3.4, 1.6 & 252 & 554 & 11 & 2 & 6.65 & Yes\\
    {\tt TM3} & 3.4, 1.8 & 215 & 343 & 9 & 2 & 8.503 & Yes\\
    {\tt TM4} & 1.8, - & N/A & N/A & 1 & 2 & N/A & No\\
    {\tt TM5} & 2.7, 1.3 & 239 & 433 & 9 & 2 & 6.875 & Yes\\
    {\tt TM6} & 4.4, 2.6 & 75 & N/A & 2 & 0.7 & N/A & No\\
    {\tt TM7} & 2.8, 2.1 & 106 & N/A & 2 & 0.8 & N/A & No\\
    {\tt TM8} & 3.1, 2.4 & 52 & N/A & 2 & 0.55 & N/A & No\\
    \enddata   
    \tablecomments{
        (1) run name,
        (2) stellar mass of the most and second-massive DMDGs (``-'' = no DMDG formed), 
        (3) line-of-sight velocity difference of the most and second massive DMDGs at $t = 2$ Gyr (``-'' = no DMDG formed), 
        (4) line-of-sight velocity difference of the most and second-most massive DMDGs obtained from orbit integration until the distance between the two most massive DMDGs that correspond to DF2 and DF4 reaches 2.1 Mpc apart (``N/A'' corresponds to the runs that do not involve orbit integration due to their lack of the number of DMDGs), 
        (5) the number of formed DMDGs with $M_{\star} > 10^{7} \msun$, 
        (6) time since the pericentric approach of the two progenitor disks when the {\sc enzo} simulation ends,
        (7) time since the pericentric approach of the two progenitor disks when the orbit integration ends, i.e., when the distance between the two most massive DMDGs is 2.1 Mpc,
        (8) whether the results from {\tt TM} (``Tailor-Made'') produce enough DMDGs to match the number of UDGs observed in the NGC1052 group.}
    \vspace{-8mm}
\end{deluxetable*}

\begin{figure*}[t]
    \centering
    \vspace{-2mm}  
    \includegraphics[width=0.85\textwidth]{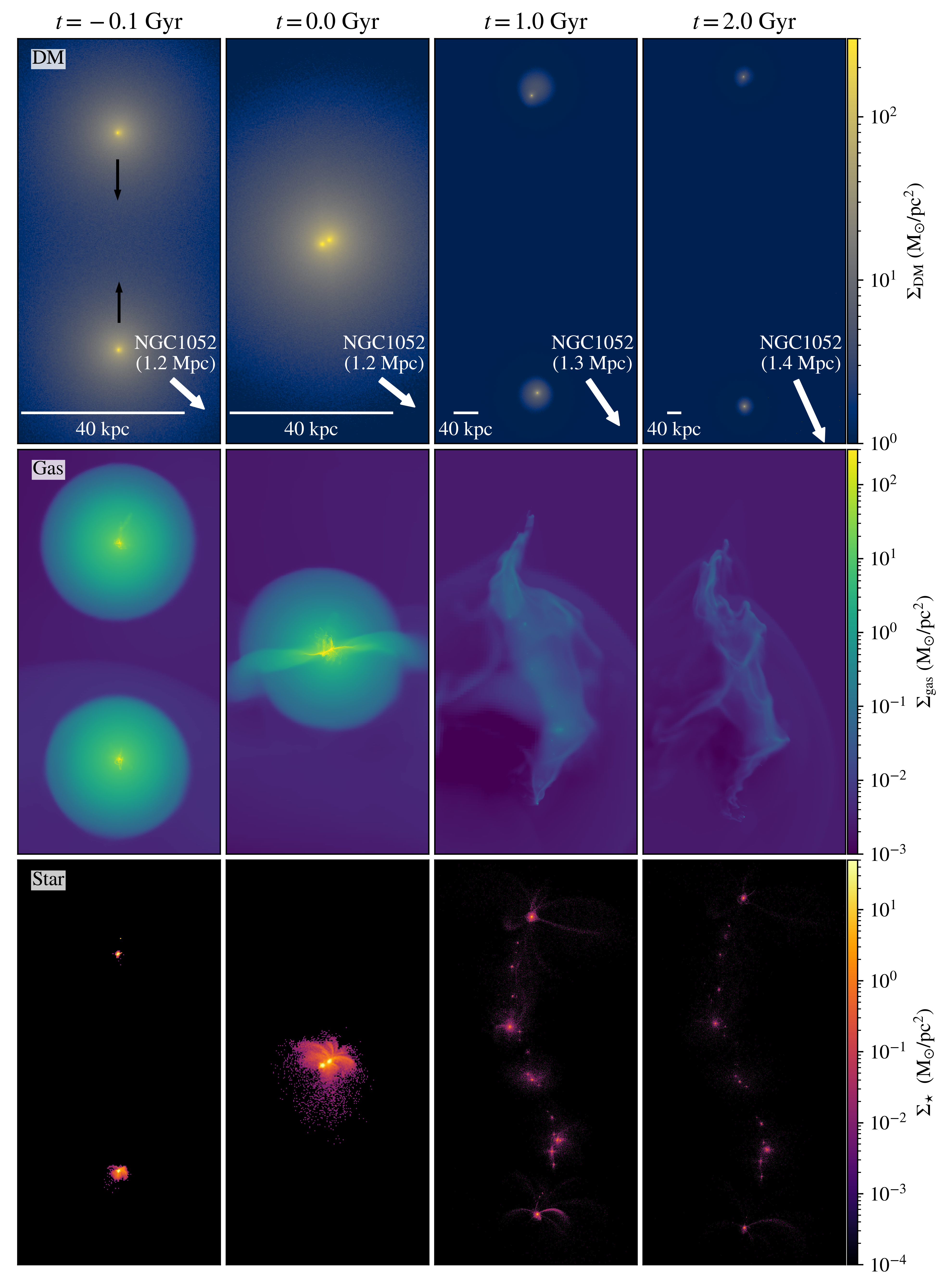}
    \caption{\label{fig:5}
        \textbf{\emph{Simulation of Mini-Bullet Satellite-Satellite Galaxy Collision from ``Tailor-Made'' Initial Conditions: Making the DMDGs/UDGs near NGC1052.}}
        Snapshots from the {\sc enzo} simulation of two colliding satellite galaxies of NGC1052 from {\tt TM1} run (``Tailor-Made'' fiducial run) initial conditions at  
        $t = -0.1$ Gyr, $t = 0$ Gyr, $t = 1$ Gyr, and $t = 2$ Gyr, centered on the two colliding satellite progenitor galaxies in the reference frame of their center of mass.
        $t = 0$ is set to the moment when the two progenitors are at a pericentric approach.
        Surface densities of dark matter (top row), gas (middle row), and stars (bottom row) in 100 kpc-depth layers are presented.
        Black arrows in the top-left panel indicate the approach of the progenitor galaxies toward each other before they collide.
        In the top row, white arrows on each plot point to the direction of the host galaxy NGC1052, centered at the distances labeled, and the length scale in each panel is indicated by the distance rulers of 40 kpc overplotted there.
        A few hundred Myr after the collision, almost no gas is left.
        At $t = 1$ Gyr and $t = 2$ Gyr, a sequence of DMDGs can be observed.
    }
    \vspace{0mm}
\end{figure*}

\begin{figure*}[t]
    \centering
    \vspace{-2mm}  
    \includegraphics[width=\textwidth]{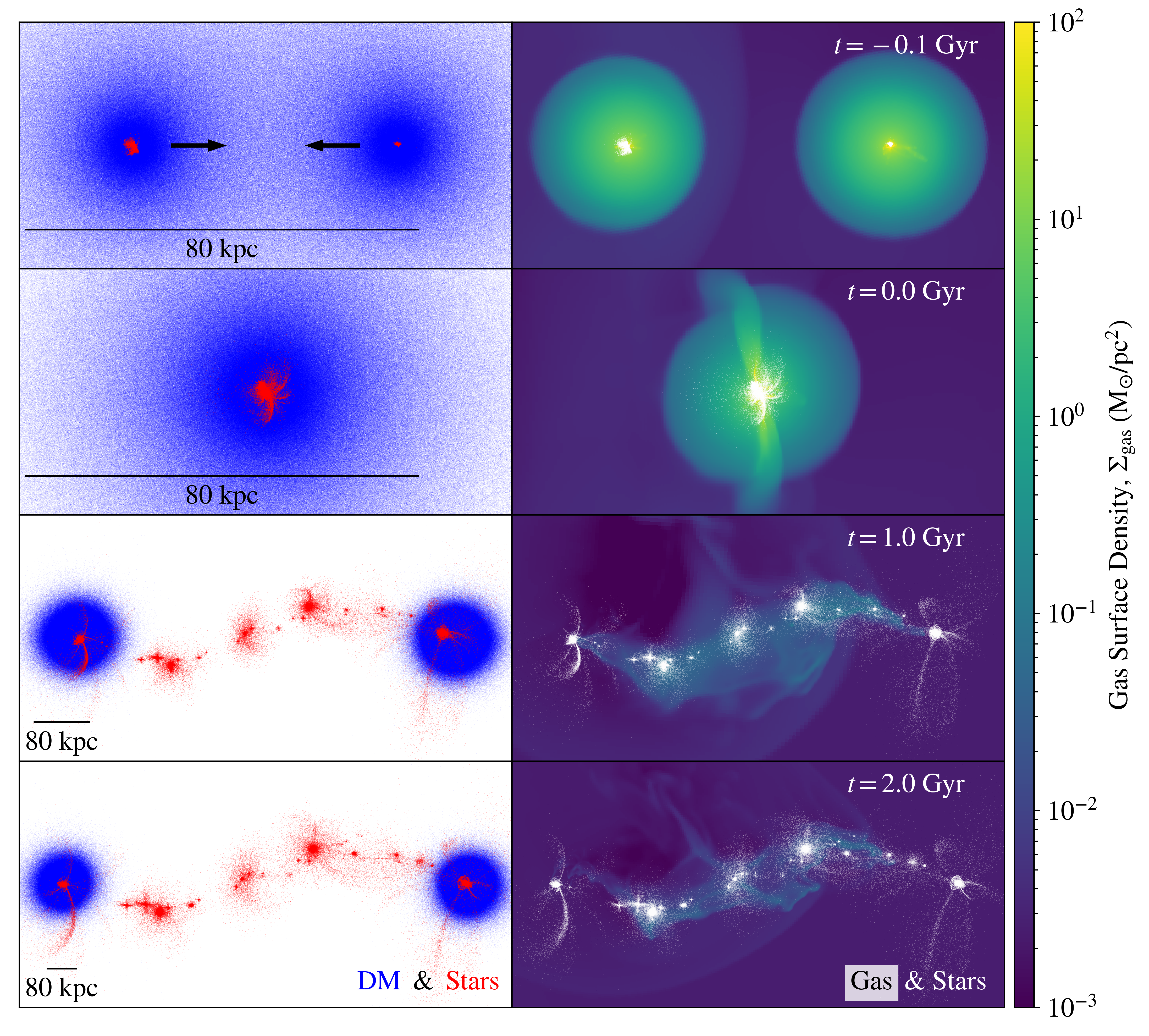}
    \caption{\label{fig:6}
        \textbf{\emph{Simulation of Mini-Bullet Satellite-Satellite Galaxy  Collision from ``Tailor-Made'' Initial Conditions: Segregating Dark Matter, Stars, and Gas.}}
        Same as Figure \ref{fig:5}, $t = -0.1$ Gyr, $t = 0$ Gyr, $t = 1$ Gyr, and $t = 2$ Gyr snapshots, but shows stars overplotted on dark matter particles and gas.
        Dark matter particles are plotted as blue (left column).
        Star particles formed after the simulation starts are overplotted red (left column) and white (right column).
        Black arrows in the top-left panel indicate the approach of the progenitor galaxies toward each other before they collide.
        Distance rulers of 80 kpc are overplotted on the left column.
        Gas is projected in 100 kpc-depth layers with a color bar showing gas surface density ($\Sigma_{\rm gas}$; right column).
    }
    \vspace{0mm}
\end{figure*}

\begin{figure*}[t]
    \centering
    \includegraphics[width=\textwidth]{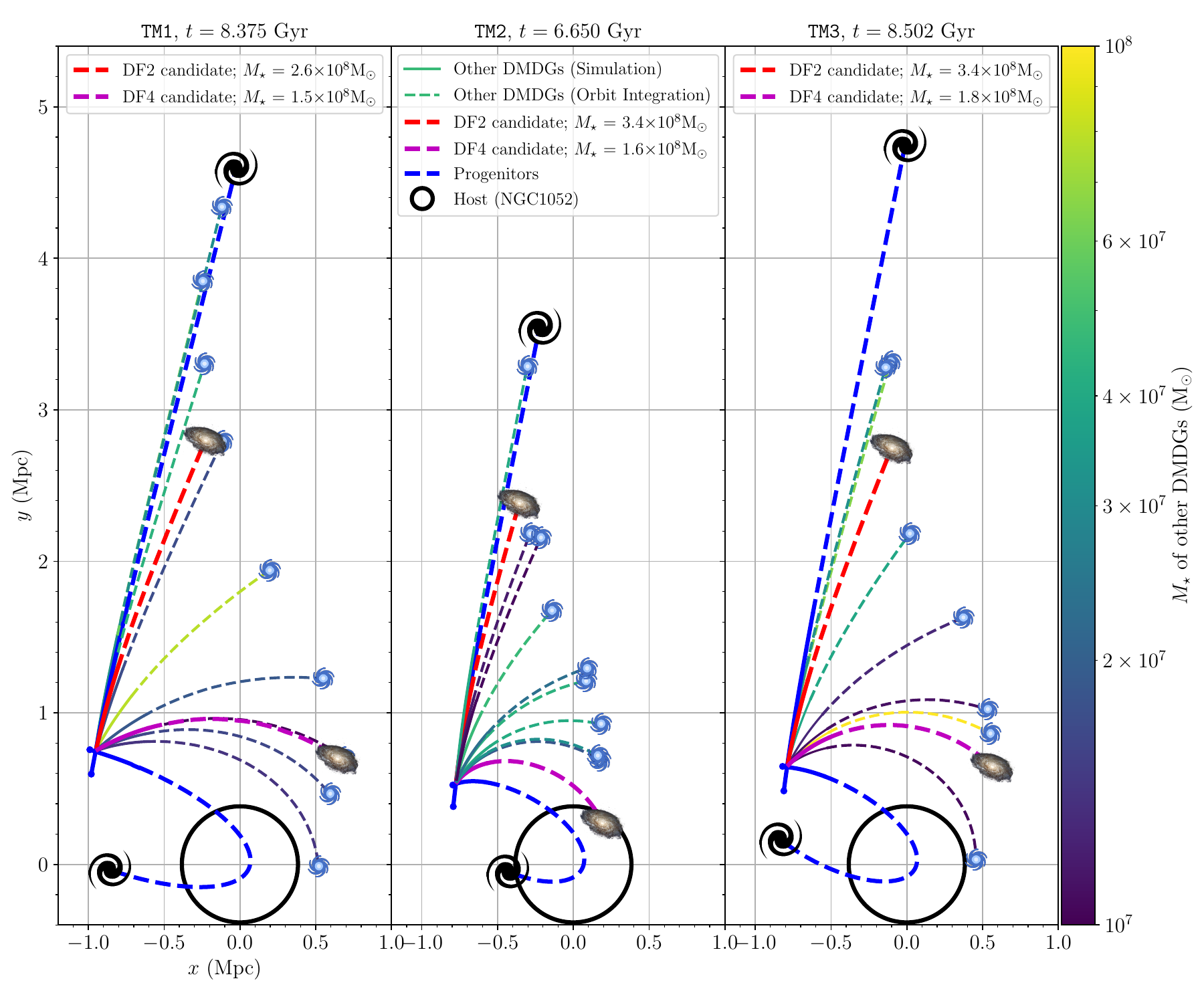}
    \caption{\label{fig:7}
        \emph{\textbf{Orbits of the progenitor galaxies and the product DMDGs.}}
        Solid lines indicate the orbits in the {\sc enzo} simulations and dashed lines are computed orbits from the {\tt Rebound} orbit integration code after the simulation end-time.
        The thick blue lines denote the trajectories of the progenitors (black galaxy icons).
        The thick red and purple lines are the orbits of the two most massive DMDGs, which correspond to DF2 and DF4, respectively (realistic galaxy icons).
        The orbits of other DMDGs are color-coded by their stellar mass indicated by the color bar on the right (small blue galaxy icons).
        The black circle indicates the virial radius of the host galaxy, NGC1052 ($R_{200} = 400$ kpc and $M_{200} = 6.124 \times 10^{12} \msun$).
        {\it Left}: Orbits in the {\tt TM1} ({\tt Tailor-made}) run for $t = -0.3 -  8.375$ Gyr.
        {\it Middle}: Orbits in the {\tt TM1} run for $t = -0.3 -  6.65$ Gyr.
        {\it Right}: Orbits in the {\tt TM1} run for $t = -0.3 -  8.502$ Gyr.
        The end time of orbit integration is when simulated DF2 and DF4 candidates are 2.1 Mpc apart in line-of-sight direction ($y$-axis).
        See Section \ref{sec:3.1} and Table \ref{tab:3} for more detail.
    }
    \vspace{0mm}
\end{figure*}

Figures \ref{fig:5} and \ref{fig:6} depicts a time sequence of the {\it Mini-bullet} collision of two satellite galaxies orbiting around a massive host, which corresponds to NGC1052 (see Section \ref{sec:2.2} and \ref{sec:2.3} for the exact host properties), and the resulting formation of spatially aligned DMDGs in the {\tt TM1} run in Table \ref{tab:2} and Table \ref{tab:3}. 
The time of the two colliding satellite galaxies at the pericentric approach is defined as $t = 0$.
At $t = 2$ Gyr, which is the last snapshot of the simulation, the progenitors are approximately 1 Mpc apart, and the sequence of DMDGs spans over 600 kpc.

The DMDGs are identified by the HOP halo finder \citep{HOP}.
We employ the boundary of each DMDG to be the tidal radius, or the Jacobi radius, the distance at which the gravitational force of the massive host becomes equivalent to the DMDG's self-gravitation, defined by \citet{Binney08}:
\begin{equation}
    \label{eq:2}
    R_{\rm tidal} = r_{\rm host}\frac{M_{\rm DMDG}}{3M_{\rm host}^{1/3}},
\end{equation} 
where $r_{\rm host}$ is the distance of the DMDG from the massive host, $M_{\rm DMDG}$ is the mass of the DMDG, and $M_{\rm host} = 6.12 \times 10^{12} \msun$ is the mass of the massive host.

The middle row of Figure \ref{fig:5} illustrates the gas distribution during and after the galaxy collision. 
Consistent with our findings in previous studies \citepalias{Shin20, Lee21}, the baryonic component of the progenitor galaxies is stripped from their dark matter halos and undergoes severe shock compression during the collision process (as shown in the second, $t = 0$ Gyr panel of the upper right panels). 
This results in a burst of star formation inside the compressed dense gas clouds \citepalias[see Figures 1 and 3 in][for more details]{Lee21}. 
Meanwhile, the stripped gas is tidally elongated, and multiple self-gravitating gas clumps form along the line connecting the two progenitors, generating stars inside these clumps. 
As a result, after approximately 1 Gyr, a number of similar stellar structures are identifiable (as shown in the bottom panel of the third column in Figure \ref{fig:5}, $t = 1$ Gyr). 
At this point, the fuel for star formation is depleted due to both the star formation burst (major) and gravitational infall to the host (minor), causing the stellar mass and population of the DMDGs to remain almost constant. 
Consequently, between $t = 1$ Gyr and $t = 2$ Gyr, the gravitational field of the host and the progenitors is the primary driver of the dynamical evolution of the DMDGs and the progenitors.

As summarized in Table \ref{tab:3}, the results from the simulations successfully satisfy two of the goals listed in Section \ref{sec:2}: the stellar mass of DF2 and DF4 and the number of product DMDGs.
We find $\sim 10$ (20; 30) self-gravitating stellar structures with $M_{\star} > 10^{7} \msun$ ($M_{\star} > 10^{6} \msun$; $M_{\star} > 10^{4} \msun$) in the simulations we analyze in detail: {\tt TM1} (Tailor-made1), {\tt TM2}, and {\tt TM3} (refer to Table \ref{tab:3} for the exact setups).
In our analysis, we restrict our focus to self-gravitating star clumps with $M_{\star} > 10^{7} \msun$ and designate them as DMDGs.
Table \ref{tab:3} also shows the velocity differences between the two most massive DMDGs, which can be compared to the observed line-of-sight velocity difference of DF2 and DF4 at the end time of the simulations ($t = 2$ Gyr for {\tt TM1}, {\tt TM2}, and {\tt TM3} run; corresponding to $\sim 6$ Gyr ago from the present), in the {\tt TM1} run and other simulations we test.
The velocity differences are smaller, but orbital motions for $\sim 6$ Gyr need to be taken into account.

Notably, the formation of DMDGs with $M_{\star} > 10^{8} \msun$ and multiple DMDGs is highly sensitive to the structural parameters of the progenitors.
Too small $r_{\rm min}$ and large $R_{\rm s,\,gas}$ result in the gathering of gas near the collision point, yielding too massive DMDGs and a small velocity difference between the most and second massive DMDGs.
Eventually, the DMDGs gravitationally pull each other and no trail of DMDGs is produced.
Furthermore, the concentration of progenitor dark matter halo also affects the DMDG formation.
More DMDGs are formed in the runs with lower concentration parameters, meaning that the strong tidal field during the {\it Mini-bullet} collision can suppress gas and newly-born stars from gathering and forming self-gravitating structures.

In order to compare our simulation results with observations of the NGC1052 group, which by design, are expected to exhibit good alignment, we further examine the orbital evolution of the DMDGs and their progenitor satellite galaxies by integrating their orbits with the {\tt Rebound} code.
Figure \ref{fig:7} displays the orbits of the product DMDGs and the progenitors, tracked in the hydrodynamic simulations from $t = -0.3$ Gyr to $t = 2$ Gyr (solid lines) and in the orbit integration from $t = 2$ Gyr to $t \sim 8$ Gyr (dashed lines).
At the latter time, the line-of-sight distance \citep[$y$-coordinates in Figure \ref{fig:4}, \ref{fig:5}, and \ref{fig:6} and $x$-coordinates in Figure 1 of][]{vD22} between the two most massive DMDGs, which correspond to DF2 and DF4, is 2.1 Mpc apart, is set to the observed amount in \citet{vD22}.
As summarized in Table \ref{tab:3}, in the {\tt TM1} ({\tt TM2}; {\tt TM3}) run, this occurs at $t = 8.375$ (6.65; 8.502) Gyr.
The line-of-sight velocity difference of the two most massive DMDGs after the orbit integration $\Delta v_{\rm DMDG,\,max2,\,now} = 324$ in the {\tt TM1} run ($554$, 343 $\kms$ in {\tt TM2}, {\tt TM3} run) is roughly in line with the observed velocity difference of DF2 and DF4, 358 $\kms$ \citep{vD22}, complying with the remaining goals of matching the velocity difference and positions of DF2 and DF4 listed in Section \ref{sec:2}.

\begin{deluxetable*}{ccccccccccccc}[t]
    \tablenum{4}
    \vspace{-1mm}
    \tablecaption{
    Positions and line-of-sight velocities of DMDGs at the end time of orbit integration, presented in Figure \ref{fig:7} from {\tt TM1}, {\tt TM2}, and {\tt TM3} runs.
    See Section \ref{sec:3.1} for details.
    \label{tab:4}
    }
    \vspace{-1mm}
    \tablewidth{0pt}
    \tablehead{
        {Run name} & {Quantity (Unit)} & {DF2 Cand.} & {DF4 Cand.} & 
        \multicolumn{7}{c}{Other aligned DMDGs}\\
        {(1)} & {(2)} & {(3)} & {(4)} & \multicolumn{7}{c}{(5)}
    }
    \startdata
    {\tt TM1} & $x$ (Mpc) & -0.23 & 0.64 
    & -0.25 & -0.23 & -0.12 & -0.12 & 0.19 & 0.52 & 0.55 & 0.59 & 0.68\\
    ($t=8.375$ Gyr) & $y$ (Mpc) & 2.80 & 0.70 
    & 3.85 & 3.30 & 4.34 & 2.79 & 1.94 & -0.01 & 1.23 & 0.46 & 0.71\\
    & $\Delta d$ (Mpc) & -0.19 & 0.06 
    & 0.08 & -0.06 & 0.34 & -0.09 & -0.03 & -0.26 & 0.11 & -0.06 & 0.10\\
    & $v_{y}$ ($\kms$) & 216.3 & -107.2 
    & 350.3 & 280.5 & 414.7 & 216.0 & 98.7 & -282.6 & -14.6 & -175.8 & -113.7\\
    & $v_{y} - v_{y,\;x{\rm \text -pred}}$ ($\kms$) & - & - 
    & 126.5 & 61.9 & 238.1 & 41.2 & 38.9 & -222.0 & 57.0 & -87.4  & 8.5\\
    & $v_{y} - v_{y,\;y{\rm \text -pred}}$ ($\kms$) & - & - 
    & -27.9 & -13.4 & -38.6 & 0.81 & 14.5 & -66.1 & 11.2 & -32.6 & -7.8\\
    \hline
    {\tt TM2} & $x$ (Mpc) & -0.36 & 0.19 
    & -0.30 & -0.29 & -0.22 & -0.14 & 0.08 & 0.09 & 0.16 & 0.18 & 0.18\\
    ($t=6.650$ Gyr) & $y$ (Mpc) & 2.38 & 0.27 
    & 3.29 & 2.18 & 2.16 & 1.68 & 1.21 & 1.30 & 0.72 & 0.69 & 0.93\\
    & $\Delta d$ (Mpc) & -0.09 & -0.10 
    & 0.19 & -0.07 & -0.01 & -0.06 & 0.04 & 0.06 & -0.02 & -0.01 & 0.06\\
    & $v_{y}$ ($\kms$) & 247.0 & -307.1 
    & 396.6 & 214.0 & 209.6 & 124.9 & 31.7 & 48.1 & -105.9 & -96.7 & -36.6\\
    & $v_{y} - v_{y,\;x{\rm \text -pred}}$ ($\kms$) & - & - 
    & 202.4 & 38.8 & 105.3 & 94.2 & 227.8 & 250.2 & 185.8 & 175.7 & 259.3\\
    & $v_{y} - v_{y,\;y{\rm \text -pred}}$ ($\kms$) & - & - 
    & -87.7 & 19.8 & 22.4 & 63.7 & 86.6 & 92.2 & 92.9 & 93.5 & 99.1\\
    \hline
    {\tt TM3} & $x$ (Mpc) & -0.101 & 0.564 
    & -0.140 & -0.109 & 0.019 & 0.370 & 0.454 & 0.532 & 0.551\\
    ($t=8.502$ Gyr) & $y$ (Mpc) & 2.75 & 0.65 
    & 3.28 & 3.32 & 2.18 & 1.63 & 0.03 & 1.02 & 0.86\\
    & $\Delta d$ (Mpc) & -0.06 & 0.02 
    & 0.05 & 0.09 & -0.09 & 0.10 & -0.25 & 0.09 & 0.06\\
    & $v_{y}$ ($\kms$) & 214.7 & -128.8 
    & 284.2 & 286.9 & 139.6 & 56.4 & -285.7 & -49.4 & -80.8\\
    & $v_{y} - v_{y,\;x{\rm \text -pred}}$ ($\kms$) & - & - 
    & 49.2 & 68.1 & -13.3 & 85.2 & -213.6 & 62.7 & 41.1\\
    & $v_{y} - v_{y,\;y{\rm \text -pred}}$ ($\kms$) & - & - 
    & -17.4 & -20.9 & 17.3 & 24.8 & -56.7 & 17.7 & 12.4\\
    \enddata
    \tablecomments{
        (1) run name,
        (2) projected positions ($x$) and line-of-sight positions ($y$) relative to the host,
        line-of-sight velocities relative to the host ($v_{y}$), 
        transverse displacement of a DMDG from the best-fitting line of the common trail of DMDGs
        ($\Delta d$; $-$ signs indicate if the $y$ coordinate of a DMDG is above ($+$) or below ($-$) the sequence in $xy$-plane),
        differences of predicted velocities $v_{y,\;x{\rm \text -pred}}$ ($v_{y,\;y{\rm \text -pred}}$) from a simple linear relationship between $x$ ($y$) positions and actual velocities $v_{y}$ of simulated DF2 and DF4 candidates $v_{y} - v_{y,\;x{\rm \text -pred}}$ ($v_{y} - v_{y,\;y{\rm \text -pred}}$),
        (3) corresponding quantities of the simulated DF2 candidate,
        (4) those of the simulated DF4 candidate,
        (5) those of other aligned DMDGs.
    }
    \vspace{-8mm}
\end{deluxetable*}

Conserving the alignment of the DMDGs at $t = 2$ Gyr depicted in Figures \ref{fig:5} and \ref{fig:6}, our orbit integration analysis reveals that the DMDGs remain spatially aligned after orbit integration, consistent with the observed peculiarity of the UDGs in the NGC1052 group \citep{Roman21, vD22}.
In several runs with different initial orbital and structural parameters of the colliding satellites, we consistently confirm that with appropriate initial conditions, multiple ($\sim 10$) aligned DMDGs with a considerable amount of stars are formed from a single {\textit {Mini-bullet}} collision.
However, note that the alignment of the DMDGs is not exactly on a line.
There are deviations in DMDG distribution from the line, at most $\sim 100$ kpc at $t=2$ Gyr and $\sim 300$ kpc at $t \sim 8$ Gyr.
Estimating these deviations in simulation could provide valuable insight for future observations in identifying which UDG is on the sequence of DMDGs and shares the common origin with DF2 and DF4, potentially providing a way to disprove or substantiate the {\textit {Mini-bullet}} scenario.

We present the spatial displacements of the DMDGs from the ``trail'' and the deviations in their line-of-sight velocities in Table \ref{tab:4}, along with their positions at the end of the orbit integration (Figure \ref{fig:7}).
The displacement from the sequence of DMDGs, denoted as $\Delta d$, is defined as the distance from the DMDG to the line $y = Ax + y_{0}$ that best fits all the DMDG positions.
Line-of-sight velocities {\bf ($v_y$)} of DMDGs other than simulated DF2 and DF4 candidates are estimated using either projected positions ($x$) or line-of-sight positions ($y$) through a simple linear relationship $v_{y,\;x{\rm \text -pred}} = Ax + v_{y,\,0}$ ($v_{y,\;y{\rm \text -pred}} = Ay + v_{y,\,0}$) derived from the positions and velocities of DF2 and DF4 candidates.
Estimating velocities based on the linear relationship derived from $x$ and $y$ results in significant differences between the estimated line-of-sight velocities $v_{y,\;x{\rm \text -pred}}$ and $v_{y,\;y{\rm \text -pred}}$, with $v_{y,\;y{\rm \text -pred}}$ being more accurate.
This is expected since the sequence of DMDGs is elongated along the $y$-axis and relative deviations in $x$ are more significant than in $y$.

\begin{figure*}[t]
    \centering
    \vspace{0mm}  
    \includegraphics[width=0.75\textwidth]{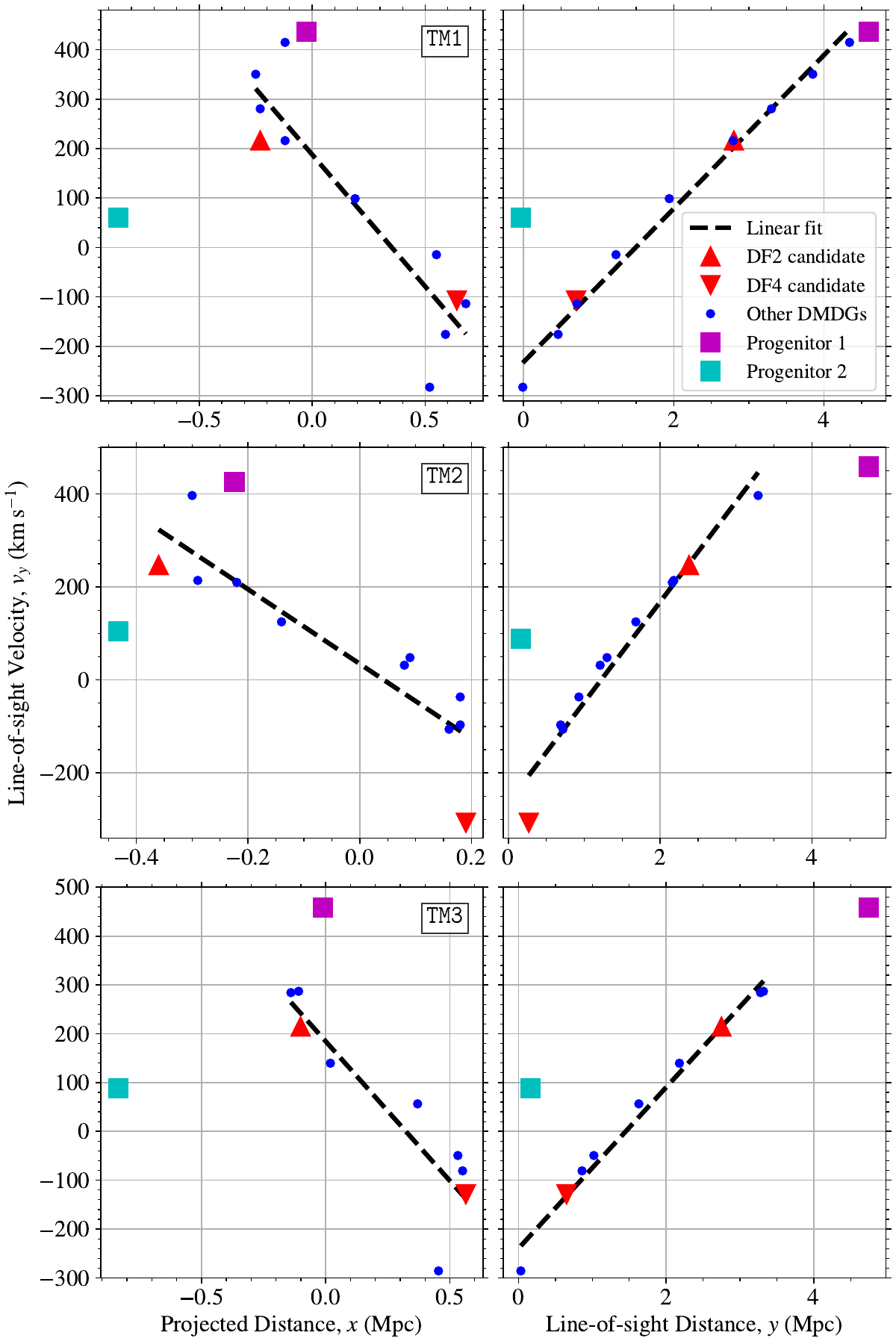}
    \caption{\label{fig:8}
        \emph{\textbf{Correlation between line-of-sight velocity and both line-of-sight and projected distance from the host galaxy.}}
        Line-of-sight velocity relative to the host galaxy ($v_y$) vs. projected distance ($x$) from the host galaxy ({\it left column}).
        Line-of-sight velocity relative to the host galaxy vs. line-of-sight distance ($y$) from the host galaxy ({\it right column}).
        Black dashed lines are linear regression fits of the correlations using the data of the product DMDGs excluding the progenitors.
        The progenitors are marked with squares, with Progenitor 1 (the one farther from the host) and Progenitor 2 (the one closer to the host).
        The correlation between $v_y$ and $y$ is tighter than that of $v_y$ and $x$.
        See Table \ref{tab:4} for the exact numbers used to plot this figure.
    }
    \vspace{0mm}
\end{figure*}

\begin{figure*}[t]
    \centering
    \vspace{0mm}  
    \includegraphics[width=\textwidth]{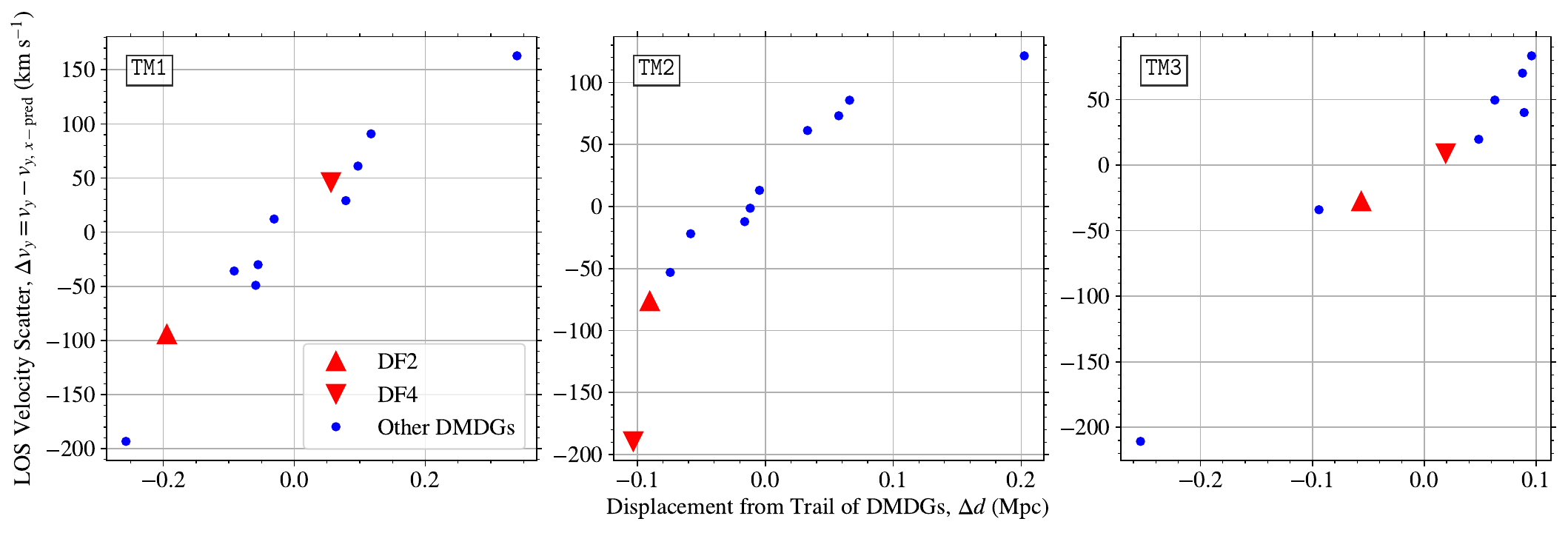}
    \caption{\label{fig:9}
        \emph{\textbf{Correlation between the predicted deviation of line-of-sight velocity and displacement of a DMDG from the trail of DMDGs.}}
        Deviation of predicted line-of-sight velocity ($v_y$) based on the linear relationship between $v_y$ and projected distance ($x$) from and its simulated value ($v_{y} - v_{y,\;x{\rm \text -pred}}$) vs. 
        transverse displacement from the line of DMDGs ($\Delta d$).
        Signs in $\Delta d$ indicate if the $y$ coordinate of a DMDG is above ($+$) or below ($-$) the sequence of the DMDGs fitted with a simple linear relationship between $x$ and $y$.
        See Table \ref{tab:4} for the exact numbers used to plot this figure.
    }
    \vspace{0mm}
\end{figure*}

In Figure \ref{fig:8}, we summarize the correlation between $v_y$ and $x$, and $v_y$ and $y$ of the product DMDGs, including simulated DF2 and DF4 candidates and the progenitors, along with linear regression fits.
We perform least-square regression linear fits to the correlations using the data of the product DMDGs shown in Table \ref{tab:4} 
(not including the data of the progenitors), overplotting the fits with dashed lines.
While both correlations follow linear relationships fairly well, the correlation between $v_y$ and $y$ is tighter than that of $v_y$ and $x$.
The scatter in the correlation between $v_y$ and $x$ is crucial in interpreting the trail of UDGs in the NGC1052 group.
For instance, even if a UDG is located at the end of the trail when projected on the sky (i.e. with the greatest projected distance from NGC1052), a different UDG, located at a smaller projected distance, can have the most extreme line-of-sight velocity.

The implications of these findings for future observations are that determining the membership of a UDG in the NGC1052 group within the sequence of DMDGs is a challenging task.
The ideal approach involves precise measurements of line-of-sight velocities, distances, and projected distances of the aligned UDGs. 
However, achieving high precision in line-of-sight distance measurements is extremely difficult even with deep imaging with a significant amount of telescope time.
Thus, the first step would involve establishing a correlation between projected distances and line-of-sight velocities.

In a study by \citet{Gannon23}, the line-of-sight velocity of DF9 was measured and found to be inconsistent with the expected linear relation between projected distances and line-of-sight velocities, if it was a member of the trail along with DF2 and DF4.  
This caused the authors to question whether DF9 was correctly identified as a member of the trail or shared a common origin with the others in the trail, whether 3D geometry in projection might alter the interpretation relative to the simple linear relationship if it was a member, or, finally, whether the idea of a common origin for all the galaxies in the trail was even correct.
As the simulations here of the {\textit {Mini-bullet}} scenario predict, however, individual DMDGs are expected to exhibit deviations in their positions and velocities from that simple linear relation, so these observations of DF9 are not inconsistent with that, so far.
By gathering data from future deep observations of multiple UDGs on the trail, it will be possible to observe more of the details of the galaxies in the apparent trail, to better compare with the {\textit {Mini-bullet}} formation scenario of the aligned UDGs in the NGC1052 group.

Another interesting prediction we make is the correlation between the predicted deviation of line-of-sight velocity and 
transverse displacement of a DMDG from the best-fitting line of the common trail of DMDGs.
This is shown in Figure \ref{fig:9}, by defining the deviations as the difference between predicted $v_{y}$ using projected positions ($x$) and the actual $v_{y}$ ($v_{y} - v_{y,\;x{\rm \text -pred}}$).

In the {\tt TM1} run, 2 Gyr after the collision, the stellar mass of the most (second) massive DMDG is $2.6 \times 10^{8} \msun$ ($1.5 \times 10^{8} \msun$), showing a good agreement to the observed stellar mass, $\sim 2.0 \times 10^{8} \msun$ for DF2 and $\sim 1.8  \times 10^{8}$ for DF4 \citep{vD18a, vD19}.
Moreover, their line-of-sight velocity difference at the end of the orbit integration ($t = 8.375$ Gyr for the {\tt TM1} run) is 324 ${\rm km\;s}^{-1}$, which is similar to the observed value of 358 ${\rm km\;s}^{-1}$ \citep[$v_{\rm DF2} = 315\;{\rm km\;s}^{-1}$ and $v_{\rm DF4} = -43\;{\rm km\;s}^{-1}$ relative to the host, NGC1052;][]{vD22}.
These results, along with those from {\tt TM2} and {\tt TM3} runs, are summarized in Table \ref{tab:4}.
The stellar masses show little variation, but the line-of-sight velocity differences can vary several times depending on the initial orbital parameters of the progenitors and the subsequent orbits of the product DMDGs. 
This is because, as the orbit of the second massive DMDG (corresponding to DF4) passes closer to the host, it experiences stronger gravity, resulting in a more negative line-of-sight velocity.

\citet{vD22} postulated that NGC1052-DF7 (hereafter DF7) and RCP32, located at the farthest end of the UDG trail, represent the remnants of the progenitor galaxies.
Notably, a follow-up study showed that the identified GCs near RCP32 exhibit stellar populations that are more akin to those of the host galaxy NGC1052, in contrast to the GCs associated with DF2 and DF4 \citep{Buzzo23}.
This finding provides support for the notion that RCP32 may be the preserved remnant of a post-collision satellite galaxy.
In Figure \ref{fig:7}, the trajectories of the progenitors in our simulations and orbit integration are illustrated.
One of the progenitors is located at the farthest from the host, roughly on the trail of DMDGs, suggesting that the progenitor might correspond to RCP32.
On the other hand, the other progenitor passes through and orbits the host halo rather than remaining in the sequence of the DMDGs and the other progenitor.
This suggests the possibility that one of the progenitor galaxies may have experienced strong tidal forces exerted by the host, potentially rendering it challenging to be detected in the present day.

\subsection{Properties of product DMDGs: stellar metallicities, ages, and sizes} \label{sec:3.2}

\begin{figure*}[t]
    \centering
    \vspace{0mm}  
    \includegraphics[width=0.95\textwidth]{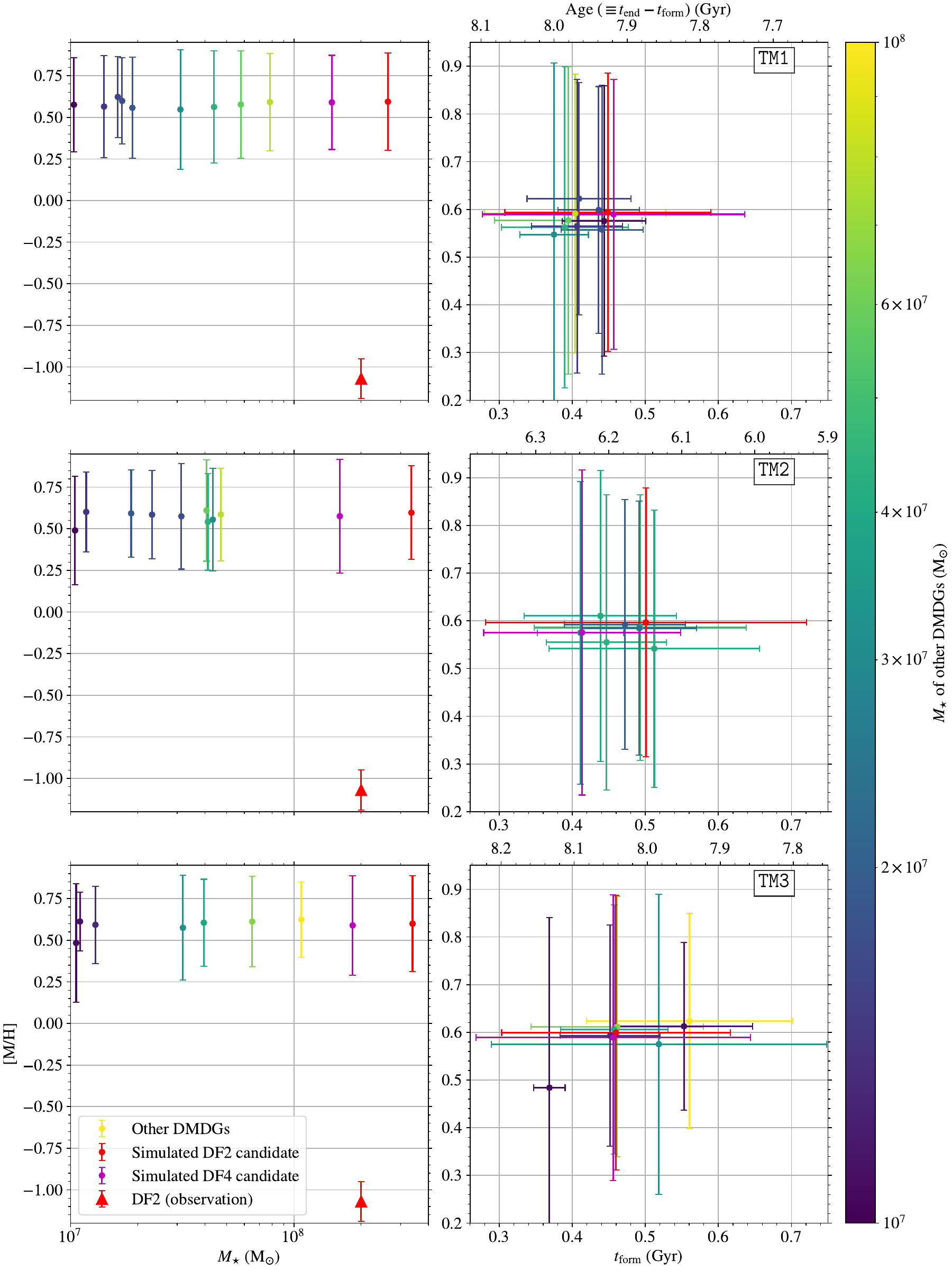}
    \caption{\label{fig:10}
        \emph{\textbf{Stellar Mass and Metallicity of the Collision-Produced DMDGs.}}
        Metallicity vs. stellar mass ({\it left column}).
        Metallicity vs. formation time (and age) ({\it right column}).
        Error bars correspond to the standard deviation of each physical quantity of the DMDG stars.
        Simulated DF2 and DF4 candidates are indicated with red and magenta dots respectively. 
    }
    \vspace{0mm}
\end{figure*}

\begin{figure*}[t]
    \centering
    \vspace{0mm}  
    \includegraphics[width=\textwidth]{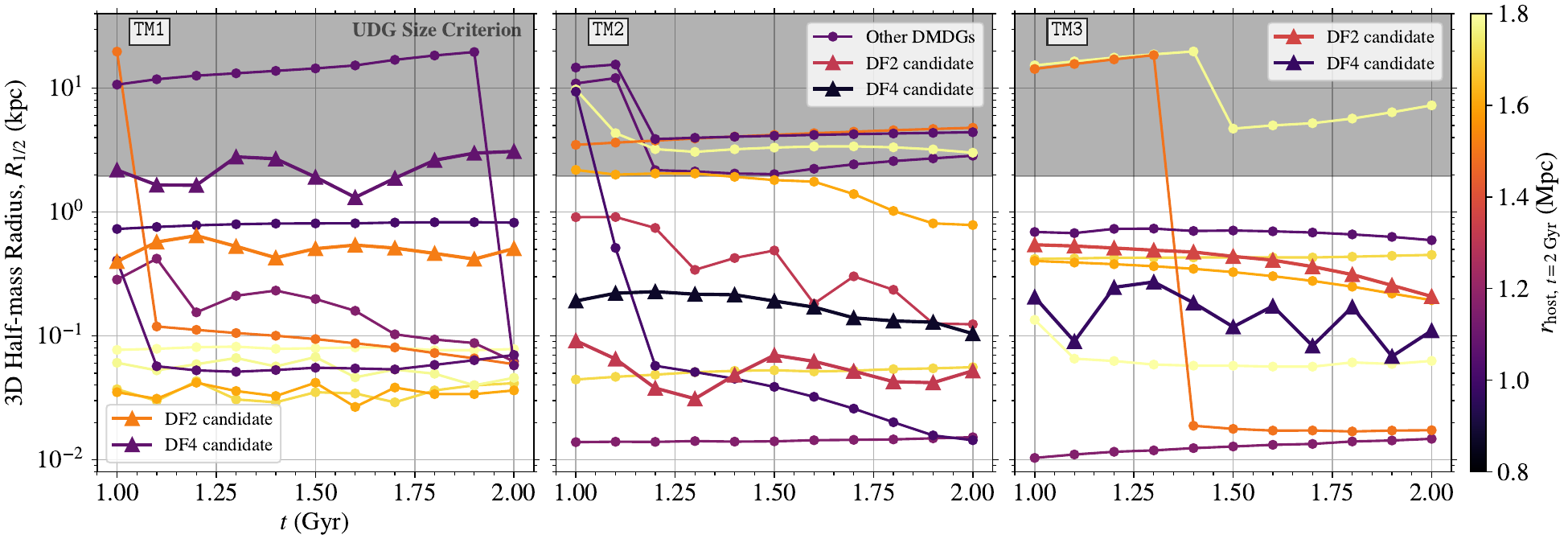}
    \caption{\label{fig:11}
        \emph{\textbf{Stellar System Sizes of the Collision-Produced DMDGs vs. UDG Size.}}
        Time evolution of three-dimensional half-mass radius $R_{1/2}$ of the product DMDGs.
        After $t=1$ Gyr, the amount of star formation of the product DMDGs is almost negligible.
        DMDGs with $M_{\star} > 10^{7} \msun$ are color-coded by their distance to the host galaxy at $t=2$ Gyr, the end of the simulations.
        Simulated DF2 and DF4 candidates are indicated with triangle markers.
        The observed stellar mass and metallicity of DF2 are marked as a red triangle on the left panels.
        Gray shade areas indicate the UDG criterion, $R_{\rm eff} \ge 1.5$ kpc ($R_{1/2} = A R_{\rm eff} =1.95$ kpc assuming $A=1.3$).
    }
    \vspace{0mm}
\end{figure*}

Now that we have confirmed that the spatial distribution, line-of-sight velocities, and stellar masses of the product DMDGs in our simulation are consistent with those observed by \citet{vD22}, we turn our attention to their detailed physical properties.
Figure \ref{fig:10} presents the stellar masses, average metallicities, and average formation time of the stars of the formed DMDGs at $t = 2$ Gyr, when the hydrodynamic simulation ends.
The two most massive DMDGs, which correspond to DF2 and DF4, are marked with red and magenta, respectively, to distinguish them from other less massive DMDGs.
On the left panel, we show the stellar masses and the average metallicities with the error bars indicating the standard deviations of the metallicities.
Overall, the average metallicities of the DMDGs exhibit a remarkable similarity with a deviation of ${\rm [M/H]} \sim 0.2$, regardless of their stellar mass.
The right panel shows the metallicities and formation time of the stars, or the age of the stars at the end of the orbit integration on the $x$-axis above, which corresponds to the observed age at the present epoch.
It is notable that massive DMDGs form stars for a longer period than other less massive DMDGs, along with the accretion of nearby gas.

Even though the initial metallicity of the gas starts with $Z = 0.1 \zsun$, the metallicities range from a few $\zsun$ to $\sim 10\zsun$, which is a lot higher than the observed metallicities of DF2 (${\rm [M/H]} = -1.07 \pm 0.12$; \citealp{Fensch19}).
We argue that this discrepancy originates from two effects: pre-enrichment of the gas in the progenitor galaxies before the collision and the stellar feedback model we employ in our simulation.
As described in Section \ref{sec:2.1}, the stellar feedback is implemented with thermal feedback of supernova.
However, simplifying various modes of stellar feedback, such as stellar winds and radiation feedback that affects nearby interstellar medium over greater distances, can result in overly efficient self-enrichment of the stars that form subsequently.
This does not mean that the metal formation is overestimated but the distribution of metal is not efficient.
Especially in the case of bursty star formation resulting from shock compression of gas due to high-velocity galaxy collision, metal distribution should be carefully modeled \citep[for example, see][for the case of GC formation in cloud-cloud collisions]{Han22}.
Including more realistic stellar feedback modes will be the subject of future studies, and our results should be interpreted as controlled simulations showing that the formed DMDGs have almost the same metallicities and ages, which can be tested in future observations.

Then in Figure \ref{fig:11}, the temporal evolutions of the 3D stellar half-mass radii ($R_{1/2}$) of the DMDGs from the {\tt TM1}, {\tt TM2}, and {\tt TM3} runs are presented.
The tracking of the DMDG is based on the star particle IDs at $t = 2$ Gyr, and the center of the DMDG is defined as the center of mass of the identified stars at that time, at which we measure $R_{1/2}$.
It is worth noting that under the assumption of spherical symmetry and density profile, $R_{1/2}$ can be converted to the effective radius, or 2D projected half-light radius, $R_{\rm eff}$, by $R_{\rm eff} = R_{1/2} / A$, where $A \sim 1.3 - 1.35$ \citep{Wolf10}.
The color-coded lines in Figure \ref{fig:11} indicate the 3D half-mass radii, with the distances from the host galaxy at $t = 2$ Gyr, $r_{{\rm host}, \, t = 2 \; {\rm Gyr}}$, determining the color scheme.

The two most massive DMDGs, denoted by larger markers (representing simulated DF2 and DF4 candidates), exhibit up to a factor of $\sim 10$ times more compact sizes compared to the observed $R_{\rm eff}$ of DF2 \citep[2.2 kpc;][]{vD18a} and DF4 \citep[1.6 kpc;][]{vD19}, except the simulated DF4 candidate in {\tt TM1} run, which has $R_{1/2} \sim 3$ kpc.
The fact that this DMDG has the size of a UDG is a novel finding and enhances the plausibility of the {\textit {Mini-bullet}} scenario.
Prior to this work, in our previous works \citepalias{Shin20, Lee21}, it has not been demonstrated that the produce can be diffuse.
The size difference in simulation runs is not negligible; the physical origin of the difference and the factors that affect the sizes of DMDGs will be discussed in detail in Section \ref{sec:4.1.1}.

Despite the suite of hydrodynamic simulations, {\tt TM1}, {\tt TM2}, and {\tt TM3}, start from nearly identical initial conditions, the sizes of the product DMDGs are notably different.
This means that the DMDG sizes are highly sensitive to various complicated physical processes involved, especially the turbulent behavior of gas driven by stellar feedback processes during the DMDG formation.
Since the simulations we perform are limited by the spatial resolution of 5 pc and simple stellar feedback physics adopted, the goal of realistically modeling the turbulent gas emerging from cloud scale and the influence of the tidal field on the distribution of gas and stars at the same time is extremely hard to achieve.
Therefore, drawing a definitive conclusion on the sizes of DMDGs in {\textit {Mini-bullet}} scenario from the simulations presented in this work is challenging.

Instead, we focus on the effect of simple physical processes including the tidal field of the host and the merging of stellar structures long after the initial burst of star formation at $t \lesssim 0.4$ Gyr.
In general, the larger and more distant the DMDG is from the host, the smaller its size becomes, and vice versa.
These results suggest that the primary factor influencing the sizes of the formed DMDGs is their susceptibility to the tidal forces exerted by the host galaxy, which is determined by a combination of the distance from the host and the mass of the DMDG.
The post-formation evolution of size is influenced by processes such as tidal interaction with the host and merging of star clumps. 
Removal of stars on the outskirts of a DMDG by tidal interaction results in a gradual decrease in size, while merging events are evident in Figure \ref{fig:11} as sudden changes in size\footnote{This is because we measure the half-mass radius to be the spatial extent of the stars in the DMDGs at the last snapshot.
The sizes of the DMDGs that experience merging are measured to be too large.
For example, in the {\tt TM1} run, one of the DMDG sizes drops just before the last snapshot, indicating that the size measured before the sudden drop is not meaningful.}.

\section{Discussion} \label{sec:4}

\subsection{Differences between the observation and simulation} \label{sec:4.1}

We have so far demonstrated that simulations starting from initial conditions designed to lead to a satellite galaxy-galaxy collision whose products match the spatial and kinematic properties and stellar masses of the NGC1052 group UDGs can indeed produce the observed trail of UDGs and their alignment, along with the dark matter deficiency of DF2 and DF4.
However, upon more detailed inspection, the simulated UDGs do not exactly reproduce all the observed characteristics of these objects.
Indeed, this is not surprising, since some details of the simulation outcome are dependent on numerical limitations and choices, such as spatial and mass resolution and the subgrid prescription for star formation and feedback physics.
In this section, we focus on how the simulated outcomes differ from observational results, especially the sizes and metallicities of the stellar components and GCs hosted by the produced galaxies.
We also discuss what needs to be taken into account for the comparison beyond our simulations by discussing other previous work on this subject.

\subsubsection{Size of DMDGs} \label{sec:4.1.1}

As presented in Section \ref{sec:3.2} and shown in Figure \ref{fig:11}, the half-mass radii of stellar components of the produced galaxies tend to be a few times smaller than the observed sizes of DF2 and DF4.
One of the factors that affect their sizes is the tidal field of the host.
Although the tidal force can reduce the size of a DMDG by removing stars from its outskirts, as appears in Figure \ref{fig:11} as a gradual decrease, heating of the stars in the central region of the galaxy can expand the DMDG \citep[see, e.g.,][]{Gnedin99, Gnedin03}.
For instance, in the context of transforming normal low-mass (dwarf) galaxies into UDGs by tidal heating, \citet{Jones21} claimed to have found possible evidence of this process in UDGs NGC2708-Dw1 and NGC5631-Dw1.
Also, while \citet{Liao19, Tremmel20, Liao19} demonstrates this in cosmological simulations, \citet{Carleton19} uses a semi-analytic model applied to dark matter-only simulation and shows the expansion of low-mass galaxies due to tidal heating.
Product DMDGs that are close to the host including the simulated DF4 candidate will experience tidal heating even after $t = 2$ Gyr, the end time of our hydrodynamic simulation, limiting the assessment of the tidal heating effect in our work.
Therefore, modeling of tidal heating of dark matter-less galaxies that occurs beyond our simulation is necessary to gauge stellar component size expansion.

On the other hand, apart from long-term size evolution on the orbits, strong supernova feedback from the formation of massive GCs during the initial star formation epoch ($t \lesssim 0.4$ Gyr) can effectively expand stars in DMDGs \citep{Trujillo-Gomez22}.
This is particularly efficient in the case of a galaxy without a co-centered dark matter halo.
Since the question of how powerful the feedback should be to expand a DMDG produced in the {\textit {Mini-bullet}} scenario into UDG needs to be studied further quantitatively, a self-consistent high-resolution simulation that resolves massive star cluster formation and its feedback is necessary to understand the expansion process.

\subsubsection{Metallicity of DMDGs} \label{sec:4.1.2}

Another difference between the simulated DMDGs from observation is that the stellar metallicities are super-solar in the simulated galaxies, whilst the observation of DF2 revealed sub-solar metallicity of $[{\rm M/H}] \sim -1$.
The overshooting of metallicity could also be observed in simulated DMDGs and star clusters in \citetalias{Lee21}, mainly caused by the simple prescription there for supernova metal ejection and an over-efficient self-enrichment -- the process that locks up metals in successive generations of star formation rather than dispersing them into the ISM--- in massive star clusters.

However, it is well known that metal dispersal within a galaxy is dependent on the computational recipe for metal mixing \citep[see e.g.,][]{Shin21}.
Furthermore, \cite{Han22} found in their radiation-hydrodynamic simulations that radiation feedback from GC stars can reduce the self-enrichment of second-generation stars by preventing immediate accretion of supernova-enriched gas and delaying subsequent star formation, which allows the released metals to mix sufficiently with interstellar gas before the next episode of star formation.
The results of this work suggest that the inclusion of radiative feedback from massive stars is crucial in studying heavy element mixing in hydrodynamic simulations.  To model the evolving chemical enrichment of gas and stars more realistically inside the star clusters formed in the galaxy-galaxy collisions simulated here, therefore, it will be necessary to add the radiative transfer of ionizing UV photons that photoionize H and He, and of optical/infrared photons that exert radiation pressure on the surrounding gas, mediated by dust.   We will consider that in future work.

\subsubsection{Survivability of globular clusters in the host tidal field} \label{sec:4.1.3}

\begin{figure*}[t]
    \centering
    \vspace{0mm}  
    \includegraphics[width=\textwidth]{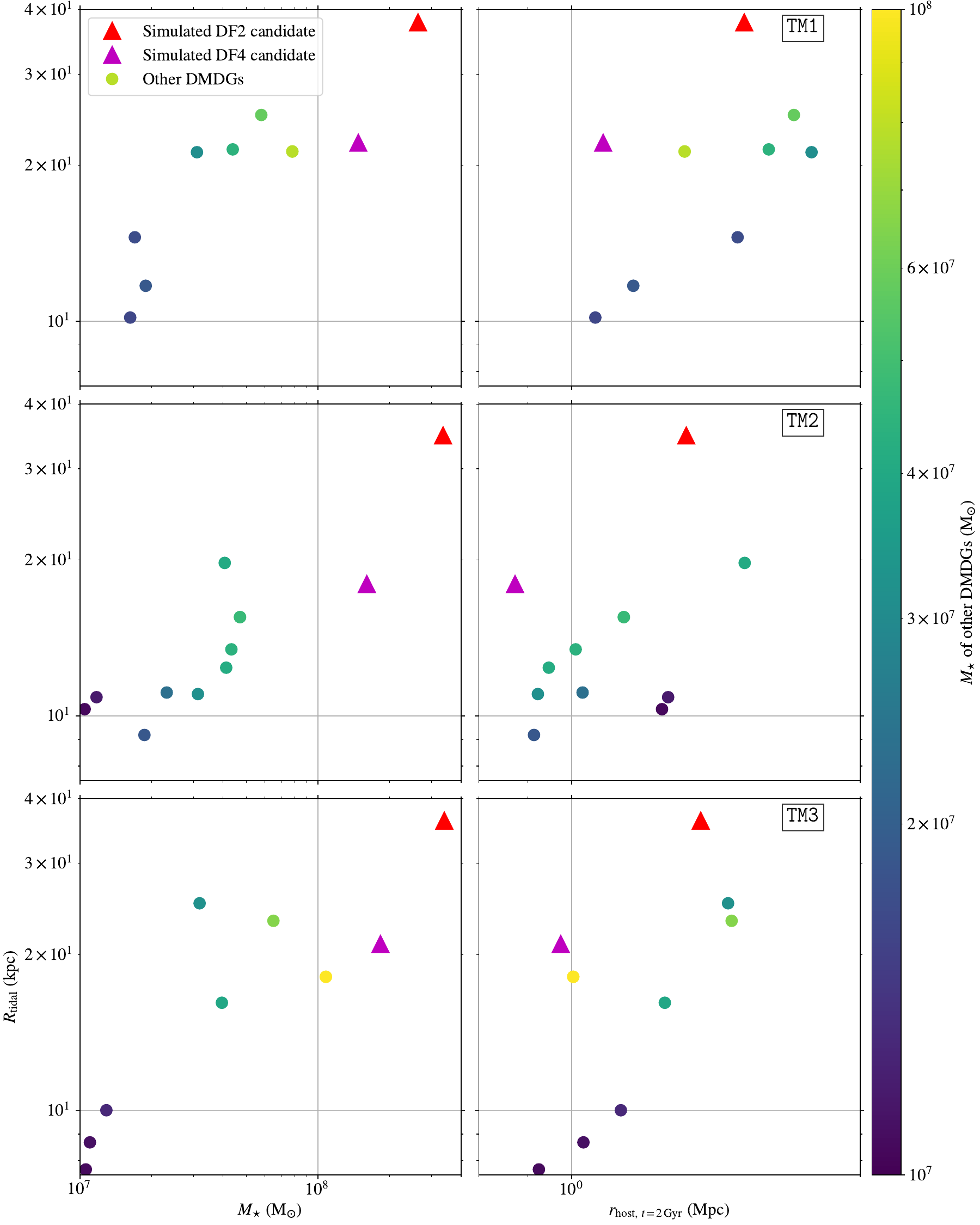}
    \caption{\label{fig:12}
        \emph{\textbf{Star-Cluster Retention Radii of the Collision-Produced DMDGs.}}
        The tidal radii $R_{\rm tidal}$ of the product DMDGs and the stellar masses $M_{\star}$ of the DMDGs ({\it left}).
        $R_{\rm tidal}$ and the distances from the host at $t = 2$ Gyr, $r_{{\rm host}, \, t = 2 \, {\rm Gyr}}$ ({\it right}).
        Simulated DF2 and DF4 candidates are indicated with red and magenta triangles respectively.
        Other DMDGs with $M_{\star} > 10^{7} \msun$ are color-coded by their stellar masses as presented in the color bar on the right.
        {\it Top}: 9 DMDGs produced in the {\tt TM1} run.
        {\it Middle}: 11 DMDGs produced in the {\tt TM2} run.
        {\it Botttom}: 9 DMDGs produced in the {\tt TM3} run.
    }
    \vspace{0mm}
\end{figure*}

\citet{Ogiya22b} argued that it is challenging for the {\textit {Mini-bullet}} scenario to explain both the number and spatial extent of massive GCs in DF2.
Considering dynamical friction, the GC distribution should have been more extended just after their formation than it is now.
Thus, the GCs are more susceptible to the strong tidal force from NGC1052, demanding a larger number of GCs at the time of the formation of DF2, meaning that most of the stellar mass in DF2 should have been in GCs, which seems unlikely.

This argument relies on the assumption that the location of the {\textit {Mini-bullet}} collision was \emph{inside} the virial radius of NGC1052, however, in order to have sufficient strength of the tidal force.
In fact, in the constrained initial conditions designed here to lead to a collision whose end-result matches the current positions and velocities of DF2 and DF4 after their orbital evolution for $\sim 8$ Gyr, that collision should have taken place much further out, at $\sim 2 \times$ the virial radius of NGC1052 (see Section \ref{sec:2.2} and Figure \ref{fig:4}).
Otherwise, if the collision point is too close to NGC1052, the alignment of DMDGs does not occur, because product DMDGs close to the host will not be part of the sequence of DMDGs because they will be accelerated by gravity and end up on the other side of the host.
Combining these, we claim that in the case of the NGC1052 group system, the {\textit {Mini-bullet}} collision should have occurred far from the host galaxy, enabling both the formation of a trail of DMDGs and extended GC distribution in the most massive DMDGs.

We support this claim with further quantitative analysis. 
Figure \ref{fig:12} displays the tidal radii $R_{\rm tidal}$, stellar masses $M_{\star}$, and the distances of the product DMDGs from the host at $t = 2$ Gyr, $r_{{\rm host}, \, t = 2 \, {\rm Gyr}}$, for each simulation run, {\tt TM1}, {\tt TM2}, and {\tt TM3}.
As a result of being farther from the host, the tidal radii $R_{\rm tidal}$ of the simulated DF2 and DF4 candidates are larger than 20 kpc, which is $> 2$ times the current spatial extent of the DF2 GCs and large enough to retain extended GCs.

\subsection{Statistical likelihood of Mini-bullet satellite-satellite collision in a large-volume $\Lambda$CDM cosmological simulation} \label{sec:4.2}

\begin{figure*}[t]
    \centering
    \vspace{0mm}  
    \includegraphics[width=\textwidth]{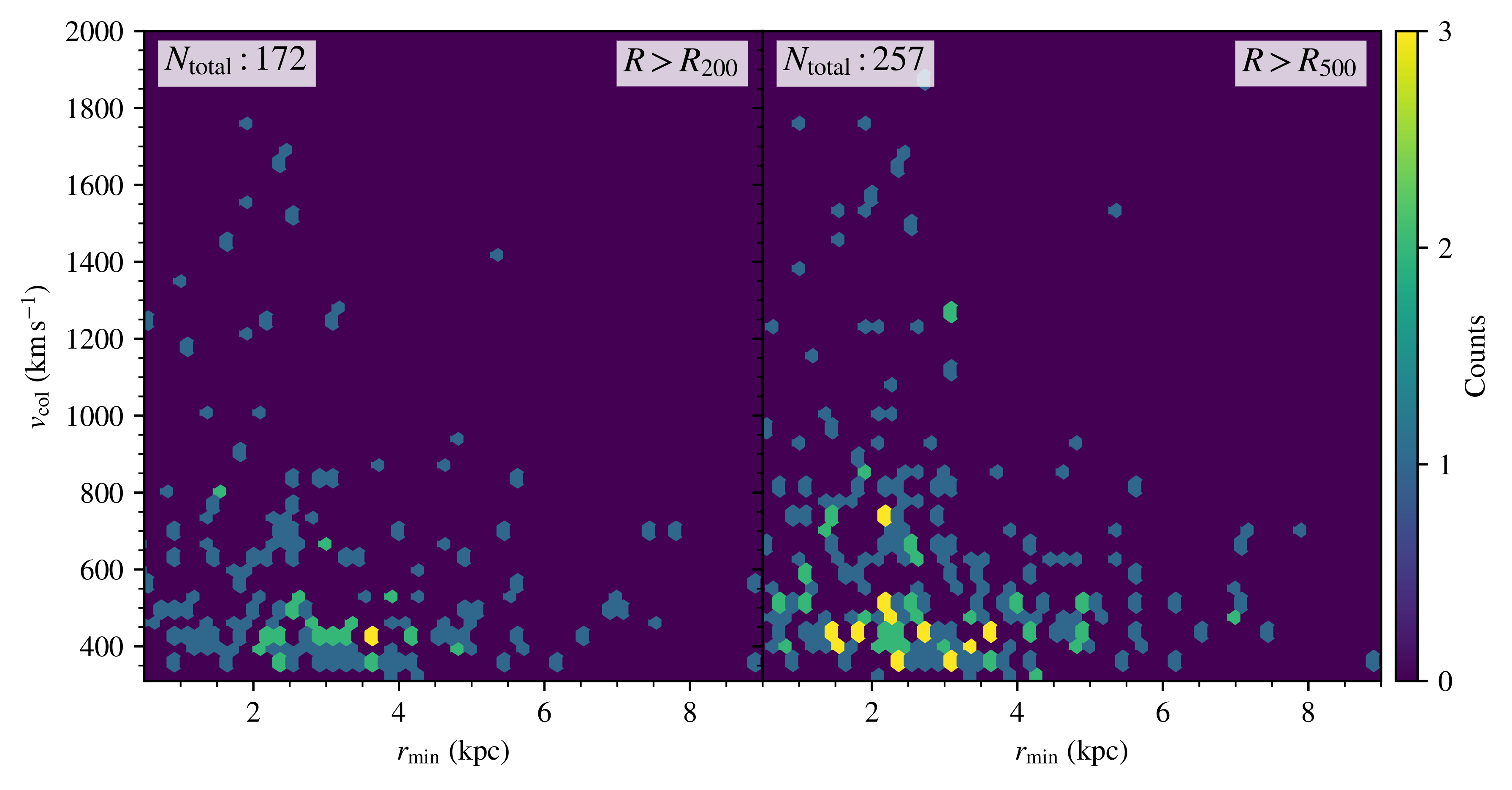}
    \caption{\label{fig:13}
        \emph{\textbf{Mini-bullet Progenitor Pairs With NGC1052-like host in a (100 cMpc)$^3$ Cosmological Simulation of $\Lambda$CDM.}}
        2D histogram of the collision pairs found in the {\sc TNG100-1} simulation volume.
        {\it Left}: collisions that occur outside of $R_{200}$ of the host halo of the collision pair.
        {\it Right}: collisions that occur outside of $R_{500}$ of the collision pair's host halo.
        The collision pairs satisfy the criteria for the relative collision velocity, pericentric distance, gas fraction, total mass ratio, gas mass ratio, and total mass of colliding galaxies and the host.
        See Section \ref{sec:4.2} for a detailed description of the criteria.
    }
    \vspace{0mm}
\end{figure*}

Quantifying the probability of the {\textit {Mini-bullet}} event is important, as such an event is expected to be rare, considering the high $v_{\rm col}$ and the small $r_{\rm min}$ of the collision.
Moreover, as discussed in Section \ref{sec:4.1.3}, the collision should occur outside of the virial radius of the host halo, possibly reducing the likelihood of its making the occurrence even further.
To address this, we will now use a large-volume cosmological simulation of galaxy and large-scale structure formation in a $\Lambda$CDM universe to study the frequency of such {\textit {Mini-bullet}} collisions in what follows\footnote{
After this work was completed, we learned of the recent paper \citet{Otaki23}, which also considered the likelihood of {\textit {Mini-bullet}} collisions a $\Lambda$CDM universe, by a semi-analytical approach.
Their focus was on collisions between subhalos \emph{inside} the virial radius of a host galaxy.   
As we have described here, however, the collisions between satellite dwarf galaxies responsible for producing DMDGs like those observed by van Dokkum in NGC1052 took place far \emph{outside} the virial radius of the host galaxy NGC1052.}.

As discussed in the previous sections, in the {\textit {Mini-bullet}} scenario, two low-mass satellite galaxies with $M_{200} \sim 10^{9 - 10} \msun$ collide with $v_{\rm col} \sim 500\;\kms$ and $r_{\rm min} \lesssim R_{\rm s,\,gas}$.
This condition is expected to be rare in the Universe, considering that the progenitor satellite galaxies collide \emph{outside} the host galaxy's virial radius (Figure \ref{fig:4}) 
with a relative velocity well in excess of that host's virial velocity but with such a small impact parameter that their centers approach within a small fraction of their own virial radii.
Moreover, as described in Section \ref{sec:2.1}, if one of the colliding satellites is unbound to the system and the other satellite is bound to the system, the likelihood of the collision will be lower.
Therefore, quantifying the statistical likelihood of the {\textit {Mini-bullet}} scenario is essential in studying how plausible the scenario is.

Towards this end, we investigate the frequency of occurrence of such satellite-satellite galaxy collision events in a large-volume cosmological N-body/hydrodynamics simulation of galaxy formation in $\Lambda$CDM, in a box 100 cMpc on a side, TNG100-1, from the simulation suite known collectively as {\sc IllustrisTNG} \citep{Pillepich18a, Pillepich18b, Nelson19, Marinacci18, Nelson18, Springel18, Naiman18}. 
{\sc IllustrisTNG} is the recent set of cosmological simulations by the {\sc Illustris} project, based on the moving-mesh code {\sc arepo} \citep{Springel2010}. 
The cosmological parameters adopted in the {\sc IllustrisTNG} are from the Planck 2015 results: $\Omega_{\Lambda,0} = 0.6911$, $\Omega_{m,0} = 0.3089$, $\Omega_{b,0} = 0.0486$, $\sigma_8 = 0.8159$, $n_s = 0.9667$, and $h = 0.6774$ \citep{Planck2016}.

TNG100-1, which has the highest resolution among the other TNG100 simulations, is composed of the $1820^3$ baryon particles with $1.4 \times 10^6 \msun$ and $1820^3$ dark matter particles with $7.5 \times 10^6 \msun$.
The particle-level data are stored in 100 snapshots from $z = 20.05$ to $z = 0$. 
In the halo catalogs, the halos identified with the friends-of-friends (FoF) algorithm \citep{Davis1985} and the subhalos identified with the {\sc subfind} algorithm \citep{Springel2001, Dolag2009} are stored. 
The {\sc IllustrisTNG} team also provides merger tree data generated by the {\sc sublink} algorithm \citep{RG2015} that traces the descendants and the progenitors of the subhalos.

In \citetalias{Shin20}, we searched for collision-induced DMDGs directly in TNG100-1 but noted that the numerical resolution of TNG100-1 was insufficient to form DMDGs.  To see DMDGs form, that simulation would have required higher mass and length resolution in both the dark matter and baryonic gas, in order to follow the shock compression of gas leading to star formation and then track the motion of those stars after they formed. 
Thus, in this section, we take a different approach, aiming to find, not the collision-induced DMDGs themselves, but rather the number of pre-collision galaxy pairs in TNG100-1 that meet the conditions established here for being possible \emph{progenitors} of the {\textit {Mini-bullet}} satellite-satellite galaxy collisions that form DMDGs.
A similar approach was also performed in \citepalias{Shin20}, with criteria applied to the two-galaxy pairs found in TNG100-1.
By contrast with our previous study \citepalias{Shin20}, however, we focus here, instead, on 3-body host-satellite-satellite systems and examine galaxy collisions in more detail by studying the location of the collision point and the gravitational boundedness of the colliding galaxies, to capture systems like those we showed here can produce the aligned galaxies of the NGC1052 system \citep{vD22}.

The exact search criteria are as follows.
Conditions {\it (\romannumeral 1)}-{\it (\romannumeral 4)} given below are applied to find the collision events that could have formed DMDGs and are based on the conditions we found to be capable of producing collision-induced DMDGs in Section \ref{sec:2.1} and our previous study \citepalias{Shin20}.

\begin{itemize}
    \item 
    {\it (\romannumeral 1)} The colliding galaxies collide $v_{\rm col} > 300\;\kms$, 
    {\it (\romannumeral 2)} the gas fraction of the galaxies, $f_{\rm gas} = M_{\rm gas}/M_{200}$, is greater than 0.05, {\it (\romannumeral 3)} the initial mass ratio of the colliding galaxies satisfies $1/3 < M_1 / M_2 < 3$, and 
    {\it (\romannumeral 4)} the initial gas mass ratio satisfies $1/3 < M_{\rm 1, \, gas} / M_{\rm 2, \, gas} < 3$. 
\end{itemize}

The following conditions {\it (\romannumeral 5)} and {\it (\romannumeral 6)} are added to find the collision events similar to the collision pairs that are studied in Section \ref{sec:3.1}.

\begin{itemize}
    \item 
    {\it (\romannumeral 5)} The total mass of each colliding progenitor galaxy, $M_1$ and $M_2$, satisfies $10^9 {\msun} < M_{1}, \, M_{2} < 10^{11} {\msun}$, and
    {\it (\romannumeral 6)} the collision is associated with a host halo and the total mass of the host halo is $M_{\rm 200,\, host} > 10^{11} \msun$. 
\end{itemize}

The following conditions are newly applied in this paper.

\begin{itemize}
    \item 
    {\it (\romannumeral 7)} The distance of the collision from the host halo is $R > R_{200}$ or $R > R_{500}$. 
    The {\textit {Mini-bullet}} satellite-satellite galaxy collision likely to produce the NGC1052 system should occur far from the host galaxy.
\end{itemize}

\begin{itemize}
    \item
    {\it (\romannumeral 8)} $r_{\rm min} < (R_{\rm eff,\,1} + R_{\rm eff,\,2})/2$, where $R_{\rm eff,\,1}$ and $R_{\rm eff,\,2}$ are the half-mass radii of the colliding galaxies. 
    To complement the temporal resolution of the TNG100-1 simulation, $r_{\rm min}$ are calculated using the {\tt Rebound} orbit integration code.
    Starting from the last snapshot before the collision, we take halos that are within 500 kpc from the colliding pair and mass of $M_{200} > 10^{9} \msun$, calculating the halos' orbits with an assumption of their being point masses.
\end{itemize}

\begin{itemize}
    \item 
    {\it (\romannumeral 9)} We select pairs in which at least one of the colliding galaxies is gravitationally unbound to the host halo, meaning the mechanical energy $E_{\rm mec} = E_{\rm K} + E_{\rm P} > 0$, where $E_{\rm K}$ is the kinetic energy and $E_{\rm P}$ the gravitational potential energy.
    $E_{\rm P}$ is calculated assuming that the host $M_{\rm 200,\,host}$ is at the center of mass of the {\sc FoF} halo.
\end{itemize}

\begin{deluxetable}{ccc}[t!]
    \tablenum{5}
    \tablecaption{
    The number of colliding satellite-satellite galaxy pairs from $z = 3$ to 0.01 found in TNG100-1 that satisfy the search criteria. 
    The numbers inside parentheses are the number of pairs from $z = 10$ to 0.01.
    See Section \ref{sec:4.2} for details.
    \label{tab:5}
    }
    \tablehead{
        \colhead{Search criteria} & \colhead{Outside of $R_{200}$} & \colhead{Outside of $R_{500}$}
        \\[-2mm]
        \colhead{(1)} & \colhead{(2)} & \colhead{(3)}
    }
    \startdata
    {\it (\romannumeral 1)} $-$ {\it (\romannumeral 7)}, $r_{\rm min} < 15$ kpc & 115 (1166) & 205 (1804)\\
    {\it (\romannumeral 1)} $-$ {\it (\romannumeral 8)} & 35 (172) & 53 (257)\\
    {\it (\romannumeral 1)} $-$ {\it (\romannumeral 9)} & 11 (36) & 14 (44)
    \enddata
    \tablecomments{
        (1) search criteria described in Section \ref{sec:4.2},
        (2) the number of collision pairs that collide outside of $R_{200}$ of the host halo, 
        (3) the number of collision pairs that collide outside of $R_{500}$ of the host halo.
    }
    \vspace{-10mm}
\end{deluxetable}

Figure \ref{fig:13} displays the distribution of the collision pairs that satisfy the search criteria {\it (\romannumeral 1)} $-$ {\it (\romannumeral 8)} in the parameter space of $v_{\rm col}$ and $r_{\rm min}$ occurring outside of $R_{200}$ or $R_{500}$ of the host halo.
With all the search criteria {\it (\romannumeral 1)} $-$ {\it (\romannumeral 9)} applied to 95 halo catalogs from $z=3$ (10) to 0.01, we find 11 (36) collision pairs happening outside of $R_{200}$ of the host halo and being unbound to the center of mass of the host halo\footnote{Note that this number is smaller than the number we got in \citetalias{Shin20}, 248.
This is mainly due to the addition of conditions {\it (\romannumeral 5)} $-$ {\it (\romannumeral 9)}.
Counting the number of collisions that happen inside of host halo $R_{200}$, this number will be much larger.}.
$z=3$ corresponds to the lookback time of 11.6 Gyr, roughly corresponds to the observed stellar age of DF2 and DF4 $+2\times$ (measurement error).
On the other hand, $z=10$ is employed as a starting point of the search range $z=10$ to 0.01 to exclude too frequent galaxy collisions at the early epoch of galaxy formation $z>10$.
Other numbers for less strict conditions are summarized in Table \ref{tab:5}.
In conclusion, at $z<3$, in the simulated $\sim (100$ Mpc)$^3$-sized Universe of TNG100-1, we attest that there are $\sim 10$ {\textit {Mini-bullet}} satellite-satellite galaxy collisions that are expected to produce a sequence of DMDGs even with the most conservative criteria.
Since these numbers are subject to the accuracy of the orbit integration, we examine and discuss this in Appendix (Section \ref{sec:a}).

\section{Conclusion and Summary} \label{sec:5}

In this paper, using gravitohydrodynamic simulations starting from the carefully designed initial conditions and orbit integrations applied to the simulations, we have demonstrated that the {\textit {Mini-bullet}} satellite-satellite galaxy collision scenario is capable of explaining the observed properties of the unusually aligned galaxies in the NGC1052 group.

Informed by results from idealized two-body hydrodynamic simulations (Section \ref{sec:2.1}), we set the initial orbital and structural parameters of the collision progenitors, one being a bound satellite galaxy and the other being an unbound satellite galaxy of a massive host galaxy that corresponds to NGC1052 (Section \ref{sec:2.2}).
We simulate satellite-satellite galaxy collision in the three-body host-satellite-satellite system with a suite of high-resolution (5pc) {\sc enzo} simulations (\ref{sec:2.3}).
The simulations are augmented with the orbit integration code {\tt Rebound} to study the orbital evolution of dark matter-deficient galaxies (DMDGs) produced in the satellite-satellite galaxy collision for $\sim 8$ Gyr (\ref{sec:3.1}).

Our main results are as follows:

\begin{itemize}
    \item 
    We demonstrate that the {\textit {Mini-bullet}} satellite-satellite galaxy collision that can reproduce the DMDGs in the NGC1052 group observed by \citet{vD22} should have taken place \emph{outside} the virial radius of the host galaxy NGC1052, i.e more than $\sim 2 \times$ the virial radius, in fact.  As a result, the globular clusters (GCs) formed during this galaxy collision are able to survive tidal disruption by the host galaxy. (Section \ref{sec:2.2} and Section \ref{sec:4.1.3})
    
    \item 
    We show that the {\textit {Mini-bullet}} satellite-satellite galaxy collision event is capable of producing a ``trail of DMDGs'' that consists of $\sim 10$ DMDGs with $M_{\star} > 10^{7} \msun$ and involves two massive DMDGs with $M_{\star} > 10^{8} \msun$ similar to the observed stellar mass of NGC1052 DF2 (DF2) and NGC1052 DF4 (DF4), whose positions and velocities are roughly in line with observation (Section \ref{sec:3.1}).

    \item 
    We find that while the positions ($\bm{x}$) and velocities ($\bm{v}$) of the aligned DMDGs approximately conform to a linear relationship, $\bm{v}=A\bm{x}+\bm{v}_{0}$, individual DMDGs can significantly deviate from that simple relation (Section \ref{sec:3.1}).
    
    \item 
    We forecast that one of the collision progenitor galaxies is likely to be located at the end of the trail of DMDGs and might be able to be confirmed to be distinct from other aligned UDGs in future observations (Section \ref{sec:3.1}).

    \item
    We investigate the stellar ages and metallicities of the product DMDGs and find that they are nearly identical within the standard deviations (Section \ref{sec:3.2}).

    \item
    We study the size evolution of the product DMDGs and conclude that the simulated DMDGs are a few times smaller than the observed size of DF2 and DF4.
    The tidal field of the host galaxy plays a major role in making a DMDG larger, but further study is needed to capture complicated physical processes during the formation of DMDGs (Section \ref{sec:3.2}).
\end{itemize}

We also discuss differences between the observations and our simulation results, concluding they are not a significant challenge to the mini-Bullet scenario (Section \ref{sec:4.1}).
The observed DMDGs, for example, are somewhat larger and more diffuse than the simulated ones.  
However, this can be explained as a possible outcome of tidal interaction with the host galaxy and nearby galaxies during the formation of DMDGs, which our simulations did not take into account. 
With higher numerical resolution, the simulations would also have resolved the formation and feedback of massive star clusters, which would serve to puff the DMDGs up (Section \ref{sec:4.1.1}).   
Inclusion of this enhanced stellar feedback would also serve to expel some of the metals before they are recycled into new stars, thereby reducing the metallicity of the simulated DMDGs, found here to exceed the observed values somewhat (Section \ref{sec:4.1.2}).
The massive GCs observed in the DMDGs DF2 and DF4 suggest that such massive star cluster formation did take place.

To quantify the statistical likelihood of the {\textit {Mini-bullet}} satellite-satellite galaxy collision event in the Universe, we inspect a large cosmological simulation TNG100-1.
We confirm that 11 galaxy collision pairs at $0.01 < z < 3$ satisfy the search criteria which we carefully choose to match the characteristics of the initial conditions that resulted in the formation of multiple aligned DMDGs similar to the NGC1052 group galaxies in the hydrodynamic simulations (Section \ref{sec:4.2}).

\section*{Acknowledgments}

Joohyun Lee would like to thank Mike Boylan-Kolchin, Karl Gebhardt, and Pawan Kumar for their valuable comments and feedback as members of his second-year graduate Astronomy research committee at The University of Texas at Austin.
The authors would also like to thank Yongseok Jo, Minyong Jung, Jorge Moreno, Boon Kiat Oh, and Pieter van Dokkum for insightful discussions. 
Shapiro and Lee are grateful for the support of NASA Grant No. 80NSSC22K1756 for a graduate fellowship issued through the Future Investigators in NASA Earth and Space Science and Technology program.
Ji-hoon Kim acknowledges support by Samsung Science and Technology Foundation under Project Number SSTF-BA1802-04. His work was also supported by the National Research Foundation of Korea (NRF) grant funded by the Korea government (MSIT) (No. 2022M3K3A1093827 and No. 2023R1A2C1003244). 
His work was also supported by the National Institute of Supercomputing and Network/Korea Institute of Science and Technology Information with supercomputing resources including technical support, grants KSC-2021-CRE-0442 and KSC-2022-CRE-0355. 
Data analysis was also performed using the computing resources from NSF XSEDE/ACCESS grant TG-AST090005 and the Texas Advanced Computing Center at the University of Texas at Austin.
The publicly available {\sc enzo} and {\sc yt} codes used in this work are the products of collaborative efforts by many independent scientists from numerous institutions around the world. 
Their commitment to open science has helped make this work possible.

\software{
	{\tt yt} \citep{YT},
        {\sc dice} \citep{Perret16},
	{\sc enzo} \citep{Bryan14, Brummel-Smith19},
	the {\sc grackle} chemistry and cooling library \citep{Smith17},
	{\tt Rebound} \citep{Rein12, Rein15},
        {\tt IPython} \citep{ipython},
	{\tt numpy} \citep{Numpy},
	{\tt scipy} \citep{Scipy},
	{\tt matplotlib} \citep{MPL}
}

\bibliographystyle{aasjournal}


\appendix

\section{Accuracy of estimating pericentric distance with orbit integration code in TNG100-1} \label{sec:a}

As briefly mentioned in Section \ref{sec:4.2}, we use the {\tt Rebound} orbit integration code to compute $r_{\rm min}$ of the colliding galaxy pairs in TNG100-1. 
To improve the accuracy of the orbit integration, we take into account the gravitational force of not only the pair of colliding galaxies but also other surrounding galaxies together if they are within 500 kpc from the colliding pair and have the total mass of $M_{200} > 10^{9} \msun$. 
Given that the galaxies are approaching with a high velocity ($\gtrsim 300\;\kms$), the orbit is computed for 200 Myr, within which the pericentric approach (collision) happens.
We assume that the galaxies behave as point mass particles and ignore the impact of the Hubble flow on the orbits of the galaxies because the orbits are calculated for a very short time (200 Myr).

Since we are interested in the distances between the colliding galaxies, not the actual position, we compare the distances between the colliding galaxies in TNG100-1 snapshots with the distances calculated using the {\tt Rebound} code to investigate the accuracy of the orbit integration.
In TNG100-1, there are only 100 snapshots so we cannot take the continuous trajectories.
We choose the snapshot right before the pericentric approach and compare the distances between the colliding galaxies at that moment. 
We study 1804 collision events in the redshift range of $0.01 < z < 10$ (the first row of Table \ref{tab:5}), in which galaxies collide with $r_{\rm min} < 15$ kpc and outside of $R_{500}$ of the host halo.
In 868 events among the 1804 collision pairs, galaxy collision happens earlier than the output time step, so we use the remaining 936 collision events for the comparison.

\begin{figure}[t]
    \centering
    \vspace{0mm}  
    \includegraphics[width=0.5\textwidth]{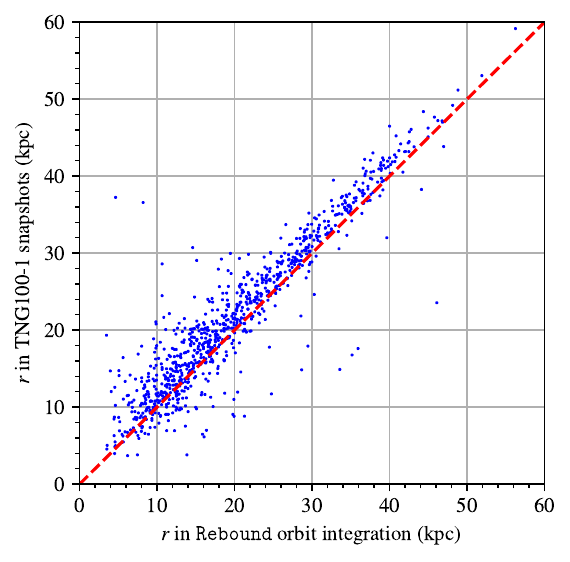}
    \caption{\label{fig:14}
        Scatter plot of the comparison of the distances between colliding galaxies in {\tt Rebound} orbit integration code and TNG100-1.
        From TNG100-1, the snapshot right before the pericentric approach is taken, and the distances are compared at that moment from {\tt Rebound} orbit integration calculation. 
        The dashed red line is the $y = x$ line.
    }
    \vspace{0mm}
\end{figure}

The results of the comparison are shown in Figure \ref{fig:14}.
The distances between colliding galaxies before the collision in {\tt Rebound} orbit integration and NG100-1 are presented.
We can see that the majority of points are located slightly above the $y = x$ line.
This means that the predicted distances of colliding galaxies in orbit integration calculation are smaller than the distances in {\sc {TNG100-1}}.
This is due to the point mass assumption of galaxies.
Also, some points are located far away from the $y = x$ line.
The main reason for this is the intervention of surrounding galaxies:
if one of the surrounding galaxies approaches one of the colliding galaxies very closely, it can disturb the orbit of the colliding galaxy.
However, the prediction from the {\tt Rebound} orbit integration is generally consistent with TNG100-1, justifying the accuracy of the orbit integration.
Thus, we argue that the existence of $\sim 10$ {\textit {Mini-bullet}} satellite-satellite galaxy collisions at $0.01<z<3$, in a $\sim 100^{3}$ Mpc volume, satisfying the search criteria {\it (\romannumeral 1)} $-$ {\it (\romannumeral 9)} described in Section \ref{sec:4.2}, is robust enough that we can believe the plausibility of the {\textit {Mini-bullet}} event in the Universe.

\end{document}